\documentclass[aps,twocolumn,prc,superscriptaddress,showpacs,nofootinbib,floatfix,amssymb,amsfonts,amsmath]{revtex4-1}
\usepackage{graphicx}
\usepackage{dcolumn}
\usepackage{bm}
\usepackage{amsmath}    

\usepackage{epsfig}
\usepackage[utf8]{inputenc}
\usepackage[T1]{fontenc}
\usepackage{color}
\usepackage{MnSymbol}

\begin{document}

\title{Parametric correlations in energy density functionals.}

\author{A.\ Taninah}
\affiliation{Department of Physics and Astronomy, Mississippi
State University, MS 39762}

\author{S.\ E.\ Agbemava}
\affiliation{Department of Physics and Astronomy, Mississippi
State University, MS 39762}
\affiliation{Ghana Atomic Energy Commission, National Nuclear
Research Institute,  P.O. Box LG80, Legon, Ghana}

\author{A.\ V.\ Afanasjev}
\affiliation{Department of Physics and Astronomy, Mississippi
State University, MS 39762}

\author{P.~ Ring}
\affiliation{Fakult{\"a}t f{\"u}r Physik, Technische Universit{\"a}t M{\"u}nchen, D-85748 Garching, Germany}

\date{\today}

\begin{abstract}
Parametric correlations are studied in several classes of covariant density functional
theories (CDFTs) using a statistical analysis in a large parameter hyperspace. In the
present manuscript, we investigate such correlations for two specific types of models,
namely, for models with density dependent meson exchange and for point coupling models.
Combined with the results obtained previously in Ref.\ \cite{AAT.19} for a non-linear
meson exchange model, these results indicate that parametric correlations exist in all
major classes of CDFTs when the functionals are fitted to the ground state properties
of finite nuclei and to nuclear matter properties. In particular, for the density dependence
in the isoscalar channel only one parameter is really independent. Accounting for these 
facts potentially allows one to reduce the number of free parameters considerably.
\end{abstract}

\pacs{21.10.Dr, 21.10.Pc,  21.10.Ft, 21.60.Jz, 21.60.Ka}

\maketitle

Since the early seventies, analogously to Coulombic quantum mechanical many-body systems, density functional theory
(DFT) has played an important role in nuclear physics. In principle, it corresponds to an exact mapping of the
complex many-body system to that of an artificial one-body system and therefore
 one with relatively small
computational costs. It is universal in the sense that the form of the energy density functional (EDF) does
not depend on the nucleus, nor on the specific region where it is applied, but only on the underlying
interaction. Thus there is only one universal functional for the Coulomb interaction
in atomic, molecular and condensed matter physics, but another one for nuclear phenomena
determined by the strong interaction and the Coulomb force.
In Coulombic systems the density functional can be derived in a microscopic way from the Coulomb force.
On the contrary in nuclear physics, because of the complexity of the nuclear
force such attempts are still in their infancy
\cite{Drut2010_PPNP64-120,SLLMR.19}.
All the successful functionals are phenomenological.
Their various forms obey the symmetries of the system, but
in the absolute majority of the cases the parameters are adjusted
to experimental data in finite nuclei and in homogeneous
nuclear matter.

Covariant density functional theories (CDFT)
\cite{VALR.05,MTZZLG.06,NVR.11,RDFNS.16,SLLMR.19}
are particularly interesting because they obey a basic symmetries of QCD.
In particular, Lorentz invariance which not only automatically includes the spin-orbit
coupling, but also puts stringent restrictions on the number of phenomenological 
parameters without loosing the good agreement with experimental data

Nonetheless, over the years, the number of phenomenological functionals has grown considerably not only
for non-relativistic Skyrme DFTs, but also for CDFTs, so that in recent years, questions have arisen about
the reliability and predictive power of such functionals \cite{DNR.14,AARR.14}. Apart from the systematic 
uncertainties which are connected with the analytic forms and the various terms in such functionals, there 
are so-called statistical uncertainties, connected with the procedures and strategies to adjust 
the various parameters to experimental data. Here we investigate whether the parameters in such CDFTs are 
independent. We search for correlations between such parameters in order to reduce their number. This will 
not only reduce the numerical efforts for determining new parameter sets, but also decrease the statistical uncertainties and, therefore, increase the predictive power of such functionals.


The Zagreb group~\cite{NV.16,NIV.17} has already tried to reduce
the number of parameters in point-coupling models with a density
dependence of exponential form, as in the functional DD-PC1~\cite{DD-PC1}.
Using the manifold boundary approximation
method (Ref.\ \cite{NIV.17}) they showed that it is possible
to reduce the number of parameters
for this functional from ten to eight without sacrificing the
quality of the reproduction of empirical data.
This method is based on the
behavior of  the penalty function in the vicinity of a minimal valley.
As designed, this method is not completely general and it still has to
be shown that it can reveal all parametric correlations in the full
parameter hyperspace.


In the present investigation we go two steps further: (i) we consider
all major classes of covariant energy density functionals (CEDFs) used
at present, and (ii) we use methods which allow us to search for such correlations
in the entire parameter hyperspace.  Our results are closely related to the
efforts of the DFT community for a microscopic derivation of EDFs and to the
search for terms which are missing in the present generation of EDFs
\cite{CDK.08,SLLMR.19}. The absence/presence of dependencies
between the parameters of the EDFs can indicate whether the terms added to the
Hamiltonian/Lagrangian have roots in physics or simply reflect additional
functional dependencies, introduced either by model approximations or by the fitting
protocol, which do not have a deeper physical context.

  There are three types of  CEDFs in the
literature, (i) those based on meson exchange with non-linear meson couplings
(NLME), (ii) those based on meson exchange with density dependent meson-nucleon
couplings (DDME), and finally (iii) those based on point coupling (PC) models containing
various zero-range interactions in the Lagrangian.
In Ref.~\cite{AAT.19} the (NLME) meson-exchange model with non-linear couplings
for the $\sigma$-mesons introduced by Boguta and Bodmer in Ref. \cite{BB.77} has been investigated
and it has been found that there is a linear correlation between the
parameters $g_2$ and $g_3$.  Within this paper we investigate parametric correlations
for the two remaining types of  CEDFs, namely, for those with density dependent meson
exchange as introduced by Typel and Wolter in Ref. \cite{TW.99}
and the point coupling model introduced by B{\"u}rvenich et al in Ref. \cite{PC-F1}.

The Lagrangians of the three different functionals can be written as:
$\mathcal{L} = \mathcal{L}_{common} + \mathcal{L}_{model-specific}$
where the $\mathcal{L}_{common}$ consist of the Lagrangian of the
free nucleons and the electromagnetic interaction.
It is identical for all three classes of functionals and is written as
\begin{eqnarray}
\mathcal{L}_{common} = \mathcal{L}^{free} + \mathcal{L}^{em}
\end{eqnarray}
with
\begin{eqnarray}
\mathcal{L}^{free} = \bar{\psi}(i\gamma_{{\mu }}\partial^\mu - m ) \psi
\label{lagrfree}
\end{eqnarray}
and
\begin{eqnarray}
\mathcal{L}^{em} = -\frac{1}{4} F^{\mu\nu} F_{\mu\nu} - e\frac{1-\tau_3}{2}\bar{\psi} \gamma^\mu\psi A_\mu.
\label{lagrem}
\end{eqnarray}

  For each model there is a specific term in the Lagrangian: for the DDME models
we have
\begin{eqnarray}
\mathcal{L}_{DDME} &=&  \frac{1}{2}(\partial{\sigma})^2 - \frac{1}{2} m_{\sigma}^2 \sigma^2 - \frac{1}{4}\Omega_{\mu\nu}\Omega^{\mu\nu} + \frac{1}{2} m_\omega^2\omega^2
\notag\\
&-&\frac{1}{4}\vec{R}_{\mu\nu}\vec{R}^{\mu\nu} + \frac{1}{2}m_\rho^2 \vec{\rho}^2-g_{\sigma}(\bar{\psi}\psi)\sigma
\notag\\
&-&g_\omega(\bar{\psi}\gamma_\mu\psi)\omega^\mu - g_\rho (\bar{\psi}\vec{\tau}\gamma_\mu\psi)\vec{\rho}^\mu
\label{lagrddme}
\end{eqnarray}
with the density dependence of the coupling constants given by
\begin{eqnarray}
g_i(\rho) &=& g_i(\rho_0) f_i(x)~~~{\rm for}~i=\sigma,\omega \\
g_\rho(\rho) &=& g_\rho(\rho_0)\exp[-a_\rho(x-1)]
\label{Eq:DD}
\end{eqnarray}
where $\rho_0$ denotes the saturation density of symmetric nuclear matter and $x=\rho/\rho_0$.
The functions $f_i(x)$ are given by the Typel-Wolter ansatz \cite{TW.99}
\begin{eqnarray}
f_i(x) = a_i \frac{1+b_i(x+d_i)}{1+c_i(x+d_i)}.
\label{Eq:Typel-Wolter}
\end{eqnarray}
Because of the five conditions $f_i(1) = 1$, $f''_i(1)=0$, and $f''_\sigma(1)=f''_\omega(1)$,
only three of the eight parameters $a_i$, $b_i$, $c_i$, and $d_i$ are independent and we finally
have the four parameters $b_\sigma$, $c_\sigma$, $c_\omega$, and $a_\rho$ characterizing
the density dependence. In addition we have the four parameters of the Lagrangian $\mathcal{L}_{DDME}$
$m_\sigma$, $g_\sigma$, $g_\omega$, and $g_\rho$. As usual  the masses of the
$\omega$- and the $\rho$-meson are kept fixed at the values $m_\omega=783$ MeV and $m_\rho=763$ MeV
\cite{DD-ME2,DD-MEdelta}. We therefore  have $N_{par}=8$ parameters in the DDME class of the models.

The NL5 class of the functionals generated in Ref.~\cite{AATG.19}  has the same model specific Lagrangian
as the DDME  class except that the coupling constants $g_\sigma$, $g_\omega$, and $g_\rho$ are constants
and there are extra terms for a non-linear $\sigma$ meson coupling. These couplings are important for the description
of surface properties of finite nuclei, especially the incompressibility~\cite{BB.77} and for nuclear
deformations~\cite{GRT.90}.
\begin{eqnarray}
\mathcal{L}_{NL5}  = \mathcal{L}_{DDME-X}   - \frac{1}{3} g_2 \sigma^3 - \frac{1}{4}g_3 \sigma^4
\label{lagrnl5}
\end{eqnarray}
For the NL5 class we have $N_{par}=6$ parameters $m_\sigma$, $g_\sigma$, $g_\omega$, $g_\rho$,
$g_2$, and $g_3$.

  The Lagrangian of the PC models contains three parts:
\\
(i) the four-fermion point coupling terms:
\begin{eqnarray}
\begin{aligned}
\mathcal{L}^{4f} = & - \frac{1}{2}\alpha_S (\bar{\psi}\psi)(\bar{\psi}\psi) - \frac{1}{2}\alpha_V (\bar{\psi}\gamma_{\mu}\psi)(\bar{\psi}\gamma^{\mu}\psi) & \\
                  & - \frac{1}{2}\alpha_{TS} (\bar{\psi}\vec{\tau}\psi)(\bar{\psi}\vec{\tau}\psi)
                  - \frac{1}{2}\alpha_{TV} (\bar{\psi}\vec{\tau}\gamma_{\mu}\psi)(\bar{\psi}\vec{\tau}\gamma^{\mu}\psi),&
\end{aligned}
\label{lagr4f}
\end{eqnarray}

(ii) the gradient terms which are important to simulate the effects of finite range:
\begin{eqnarray}
\begin{aligned}
\mathcal{L}^{der} = &- \frac{1}{2}\delta_S \partial_{\nu}{(\bar{\psi}\psi)}\partial^{\nu}{(\bar{\psi}\psi)} & \\
& - \frac{1}{2}\delta_V \partial_{\nu}{(\bar{\psi}\gamma_{\mu}\psi)}\partial^{\nu}{(\bar{\psi}\gamma^{\mu}\psi)}& \\
&- \frac{1}{2}\delta_{TS} \partial_{\nu}{(\bar{\psi}\vec{\tau}\psi)}\partial^{\nu}{(\bar{\psi}\vec{\tau}\psi)}&\\
& - \frac{1}{2}\delta_{TV} \partial_{\nu}{(\bar{\psi}\vec{\tau}\gamma_{\mu}\psi)}\partial^{\nu}{(\bar{\psi}\vec{\tau}\gamma^{\mu}\psi)},&
\end{aligned}
\label{lagr4f}
\end{eqnarray}
(iii) The higher order terms which are responsible for 
the effects of medium dependence
%
\begin{eqnarray}
\begin{aligned}
\mathcal{L}^{hot}  = & - \frac{1}{3}\beta_S (\bar{\psi}\psi)^3 - \frac{1}{4}\gamma_S (\bar{\psi}\psi)^4&   \\
                    & - \frac{1}{4}\gamma_V[(\bar{\psi}\gamma_{\mu}\psi)(\bar{\psi}\gamma^{\mu}\psi)]^2. &
\end{aligned}
\label{lagrhot}
\end{eqnarray}
For the PC models we have $N_{par}=9$ parameters $\alpha_S$, $\alpha_V$, $\alpha_{TV}$, $\delta_S$, $\delta_V$, $\delta_{TV}$, $\beta_S$, $\gamma_S$, $\gamma_V$. In these calculations we neglect the scalar-isovector channel, i.e. we use $\alpha_{TS}=\delta_{TS}=0$, because it has been shown in Ref.~\cite{DD-MEdelta}, that the information on masses and radii in finite nuclei does not allow one to distinguish the effects of the two isovector mesons $\delta$ and $\rho$. The particular realizations of the DDME and PC
models used in the present manuscript, which depend on the details of fitting protocol, are labeled here as DDME-X and PC-X, respectively.

  In order to determine the $N=N_{par}$ parameters ${\bf p}=(p_1, p_2, ..., p_N)$ of our model we adjust them
to a set of $N_{data}$ data points. These data points belong to $N_{type}$ of different types and for each type,
labeled by $i$, there are $n_i$  data points of the same type, which means
\begin{eqnarray}
N_{data}= \sum_{i=1}^{N_{type}}n_i.
\end{eqnarray}
The experimental value of the physical observable $j$ of type $i$ is given by $O^{exp}_{i,j}$ and the corresponding value calculated with our model and the parameter set $\bf p$ is $O_{i,j}({\bf p})$. Adopting for each of the physical observables an error $\Delta O_{i,j}$ which, for the functionals under study, are summarized in Table 1
of the supplementary material\footnote{Note that contrary to previous studies all minimizations
of the functionals are performed within the Relativistic Hartree-Bogoliubov
(RHB)  framework with separable pairing of Ref.\ \cite{TMR.09} scaled
according to Ref.\ \cite{AARR.14}. For the rest, the fitting protocols of the
DDME-X and PC-X functionals are identical to the fitting protocols
of the functionals DD-ME2 and PCPK-1 functionals defined in Refs.\ \cite{DD-ME2,PC-PK1}.
In a similar fashion, the fitting protocol of the NL5(E) functional is very
similar to the one of the NL3* one (see Ref.\ \cite{AAT.19} for details).
The optimal DDME-X and PC-X functionals (see Tables II and III of the
supplementary material) are defined by the simulating annealing method
and by numerous applications of the simplex method (see Ref.\ \cite{NumRec}
for a description of the method). Note that DDME-X and PC-X have better penalty
function as compared to the original parameter sets DD-ME2 and PC-PK1.},
we introduce for each parameter set $\bf p$ the penalty function
\begin{eqnarray}
\chi^2({\bf p})=\sum_{i=1}^{N_{type}} \sum_{j=1}^{n_i} \left( \frac{O_{i,j}({\bf p})-O_{i,j}^{exp}}
{\Delta O_{i,j}} \right)^2
\label{Eq:1}
\end{eqnarray}
and the optimal parametrization is found for the parameter set ${\bf p}_0$ corresponding to the  minimum of the penalty function $\chi^2({\bf p}_0)$. We measure the overall quality of the calculated results by defining the {\it normalized objective function}
\begin{eqnarray}
\chi^2_{norm}({\bf p})=\frac{1}{s}\chi^2({\bf p})
\label{Ksi}
\end{eqnarray}
where the normalization factor
\begin{eqnarray}
s=\frac{\chi^2({\bf p}_0)}{N_{data}-N_{par}}
\end{eqnarray}
is a global scale factor, called the Birge factor \cite{Birge.32} and
defined at the optimal parametrization.
This leads to an average $\chi^2 ({\bf p}_0)$ per degree of freedom
equal to one \cite{DNR.14}.

\begin{figure*}[htb]
\centering
\includegraphics[angle=-90,width=5.9cm]{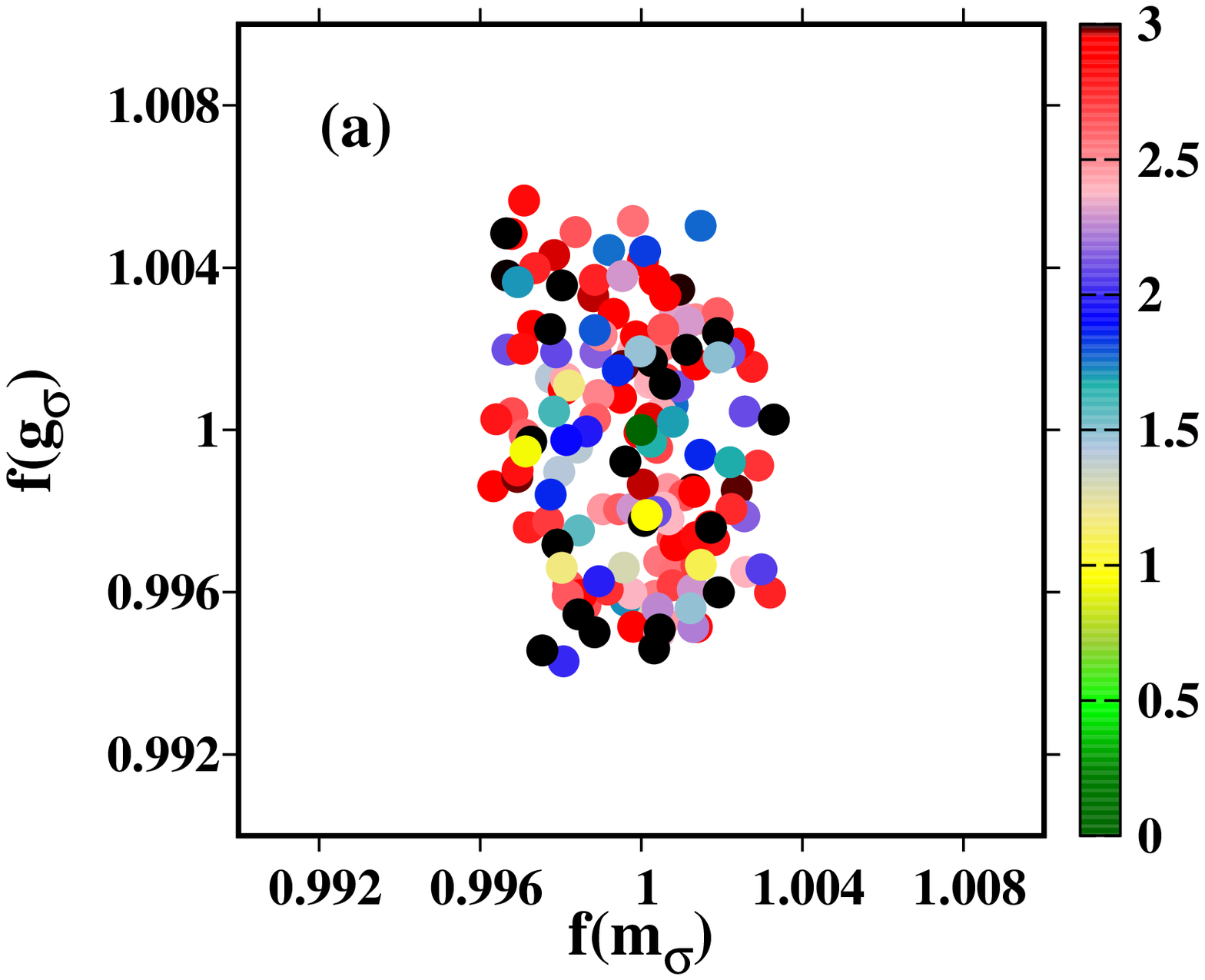}
\includegraphics[angle=-90,width=5.9cm]{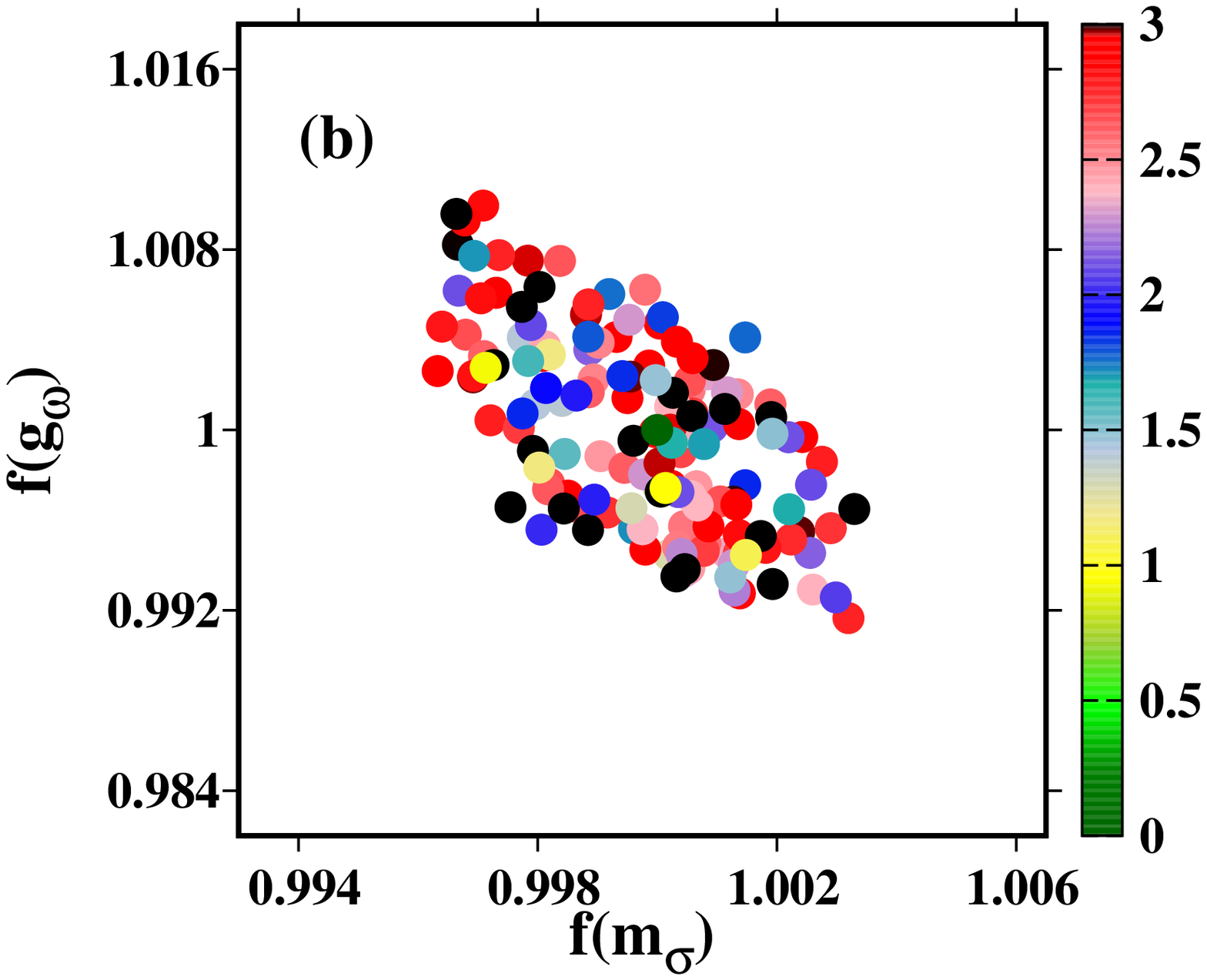}
\includegraphics[angle=-90,width=5.9cm]{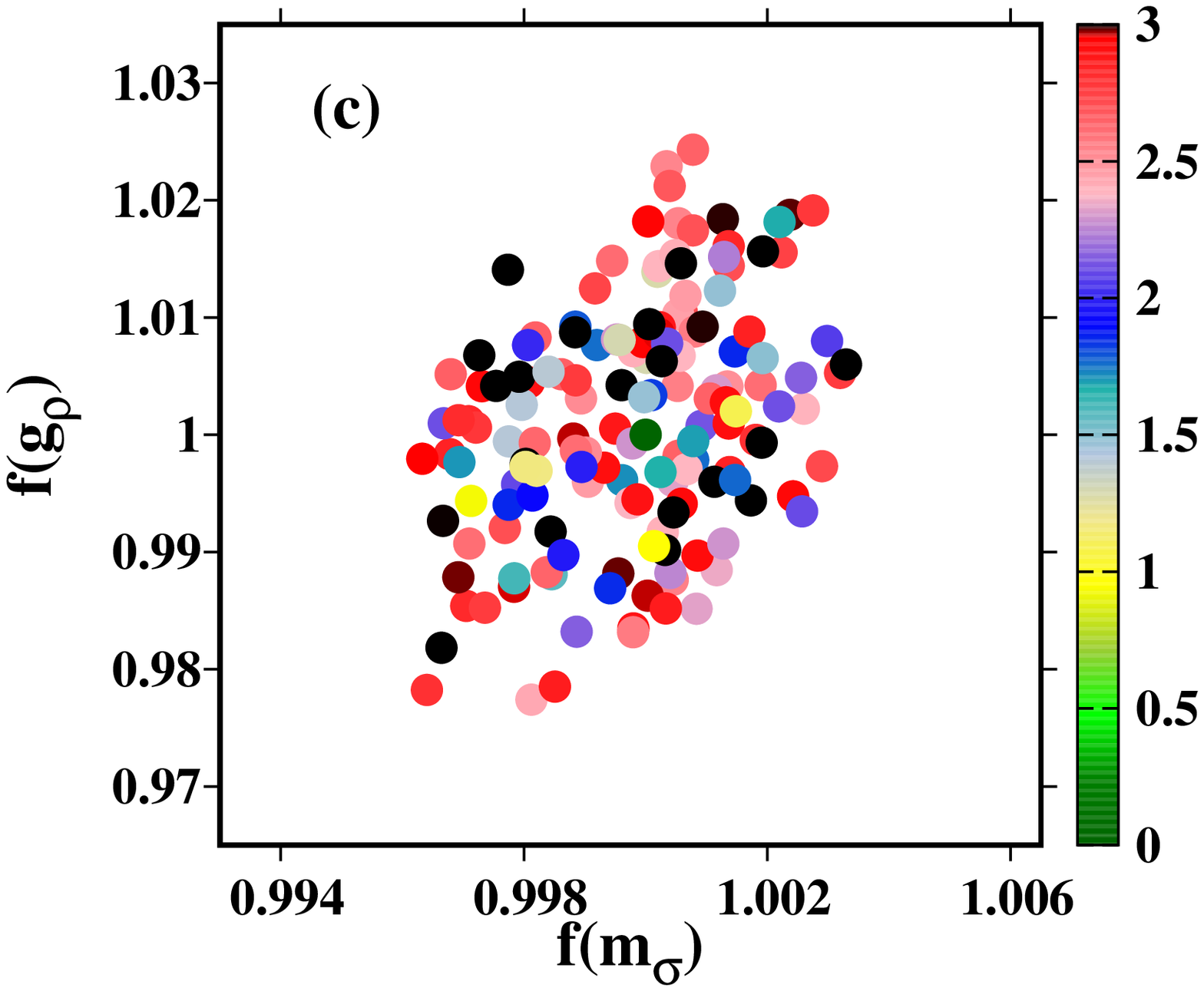}
\includegraphics[angle=-90,width=5.9cm]{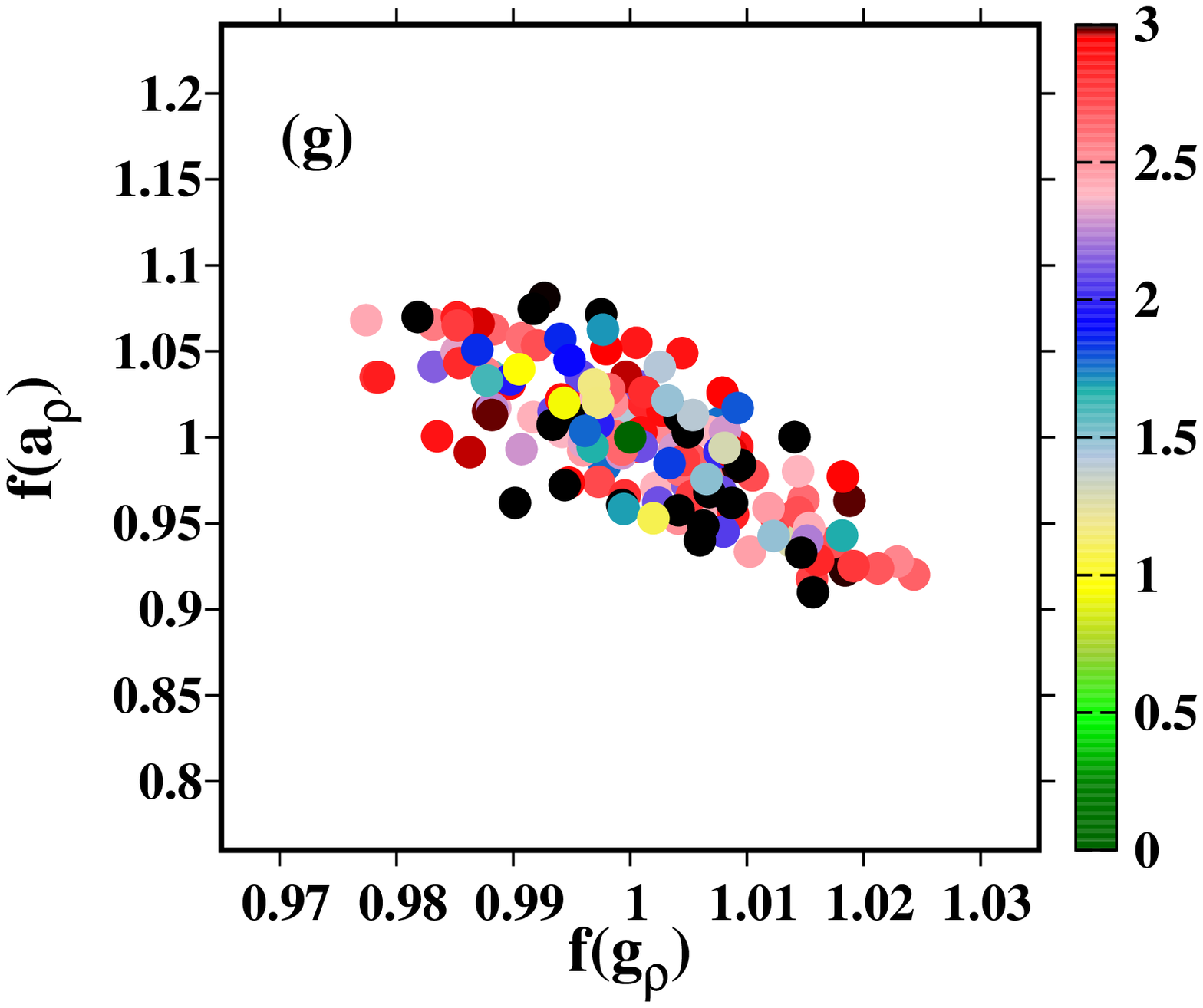}
\includegraphics[angle=-90,width=5.9cm]{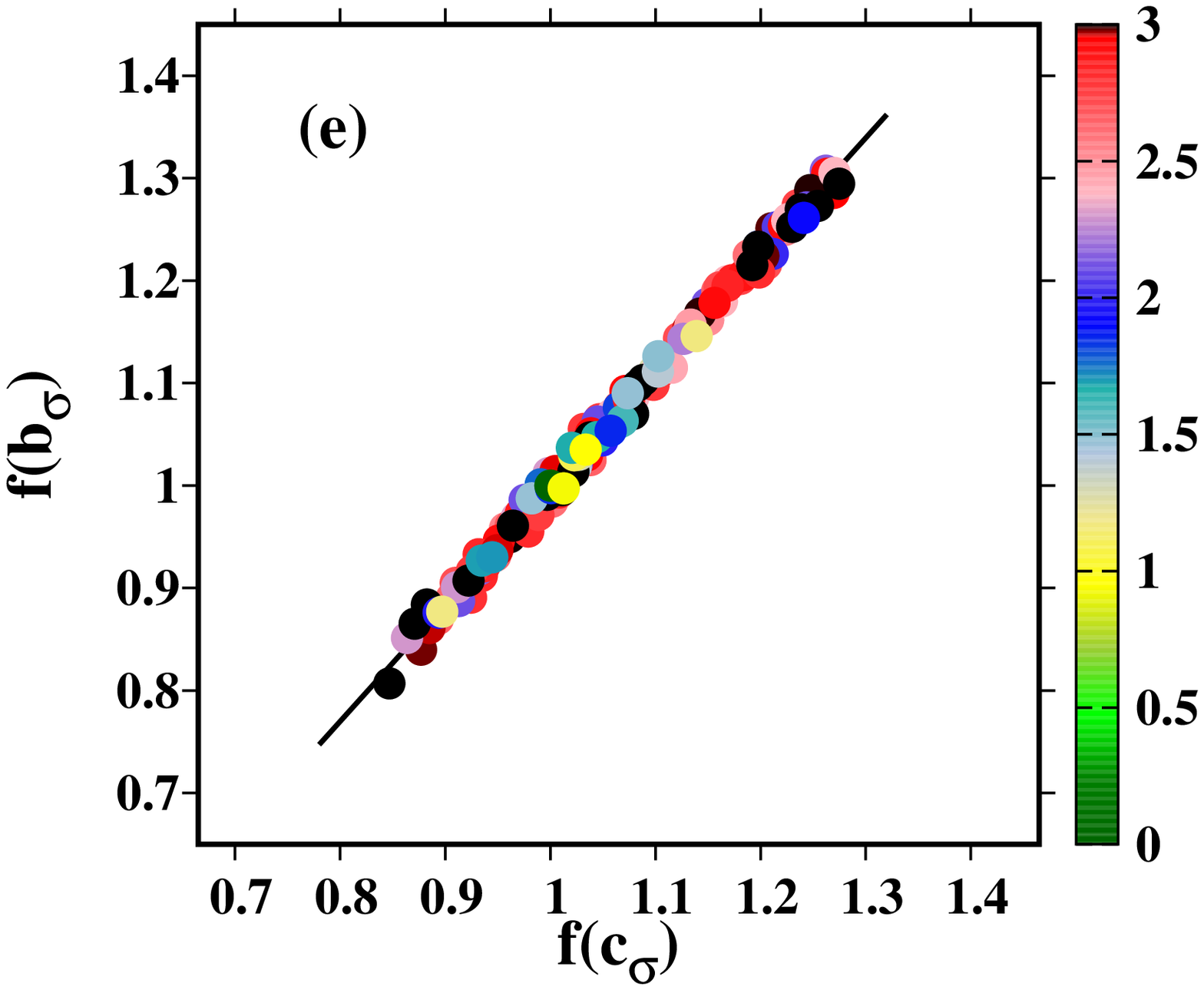}
\includegraphics[angle=-90,width=5.9cm]{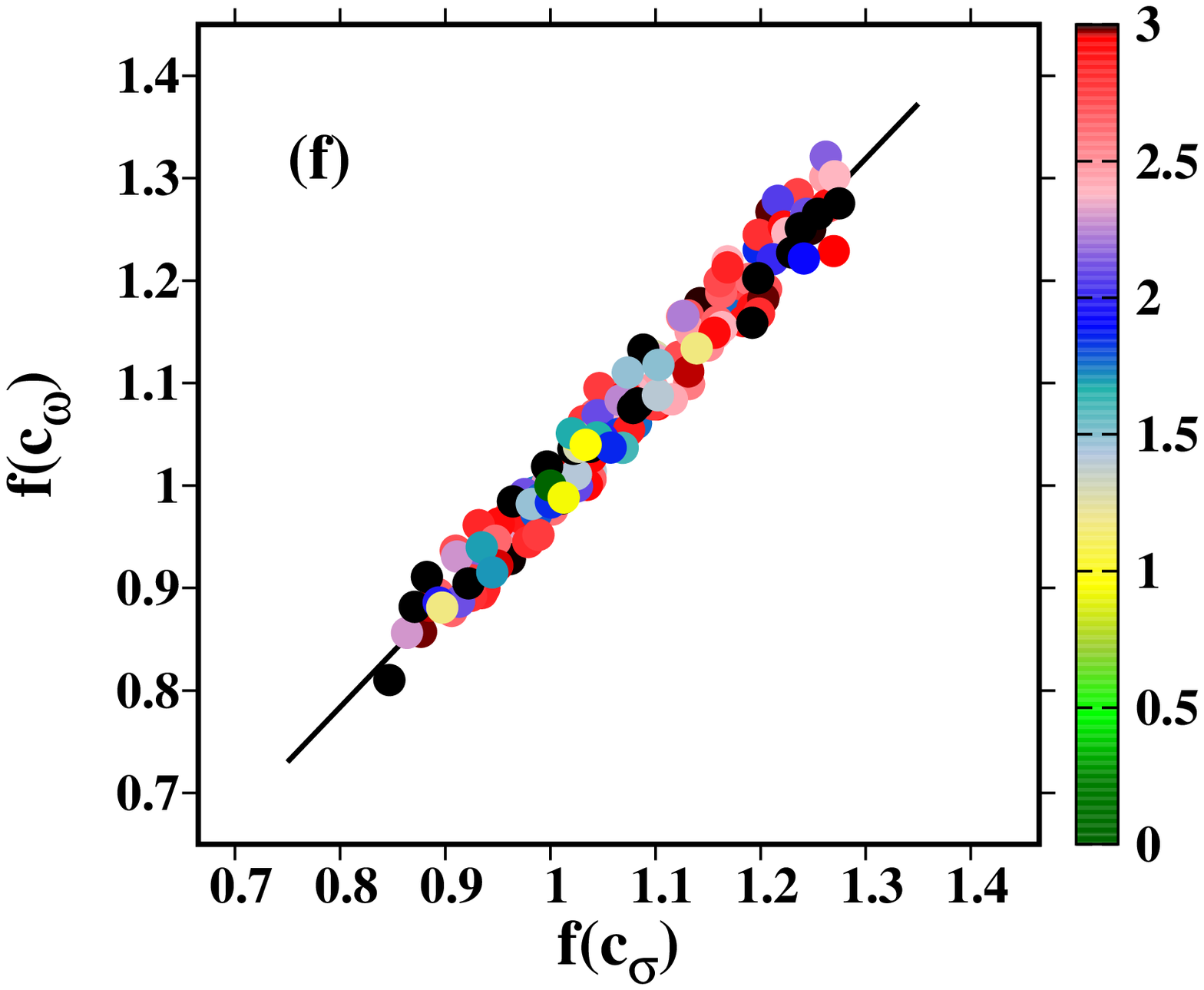}
\caption{Two-dimensional projections of the distribution of the functional variations in the 8-dimensional
parameter hyperspace of the DDME-X functional. The colors indicate the $\Delta \chi^2$ value of
the $\chi^2_{norm}({\bf p})$
of the functional variation where the latter is expressed as
$\chi^2_{norm}({\bf p})=\chi^2_{norm}({\bf p}_0) + \Delta \chi^2$. A color map
is used for the functional variations with maximum value of  $\Delta \chi^2$
equal to $\Delta \chi^2_{max} = 3.0$; there are 200 such
variations.
The optimal functional is located at the intersection of the lines $f(p_k)=1.0$
and $f(p_j)=1.0$. The solid lines in panels (e) and (f) display the parametric
correlations between the respective parameters.
}
\label{fig-DDME-X-stat}
\end{figure*}

\begin{figure*}[htb]
\centering
\includegraphics[angle=-90,width=4.0cm]{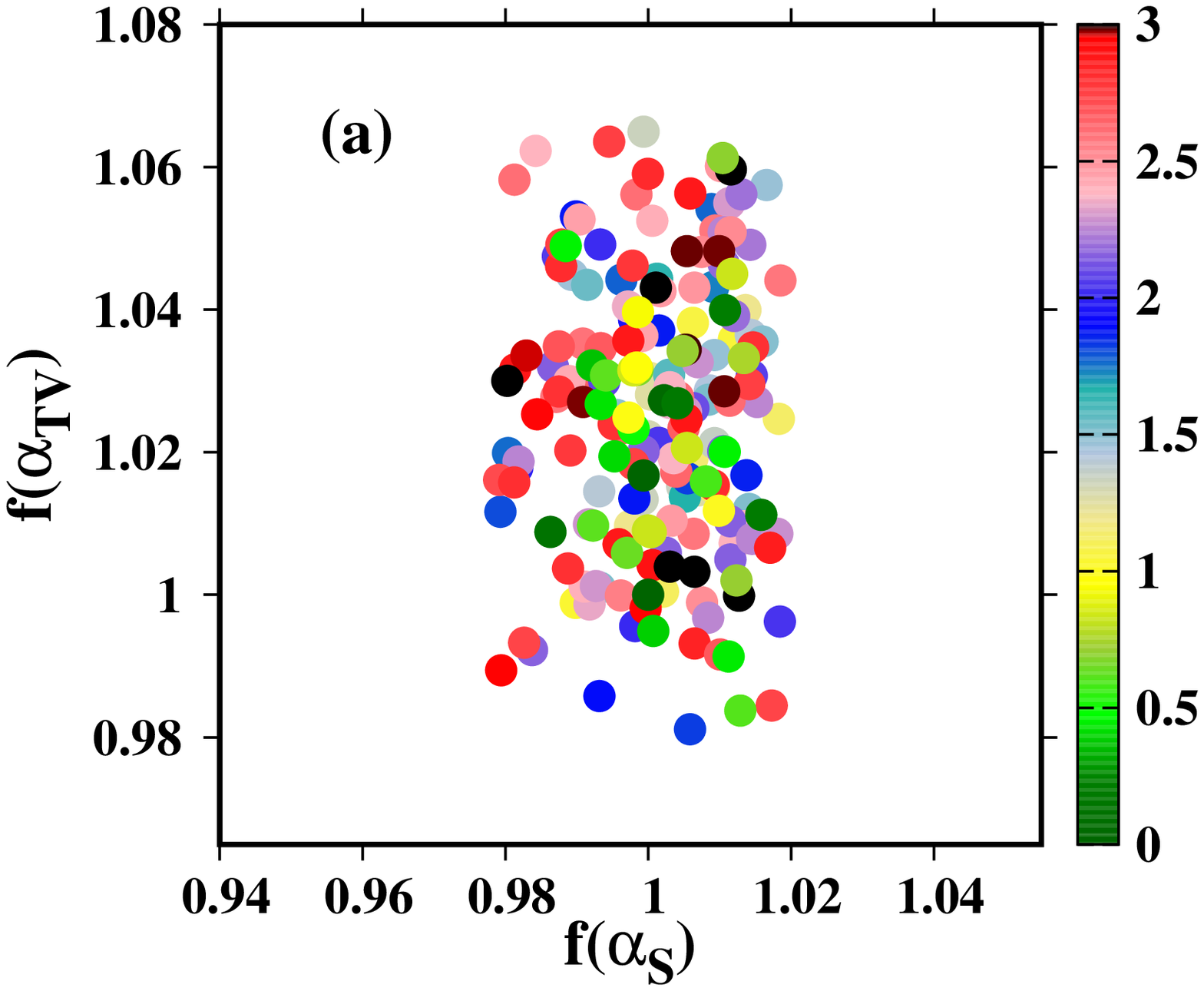}
\includegraphics[angle=-90,width=4.0cm]{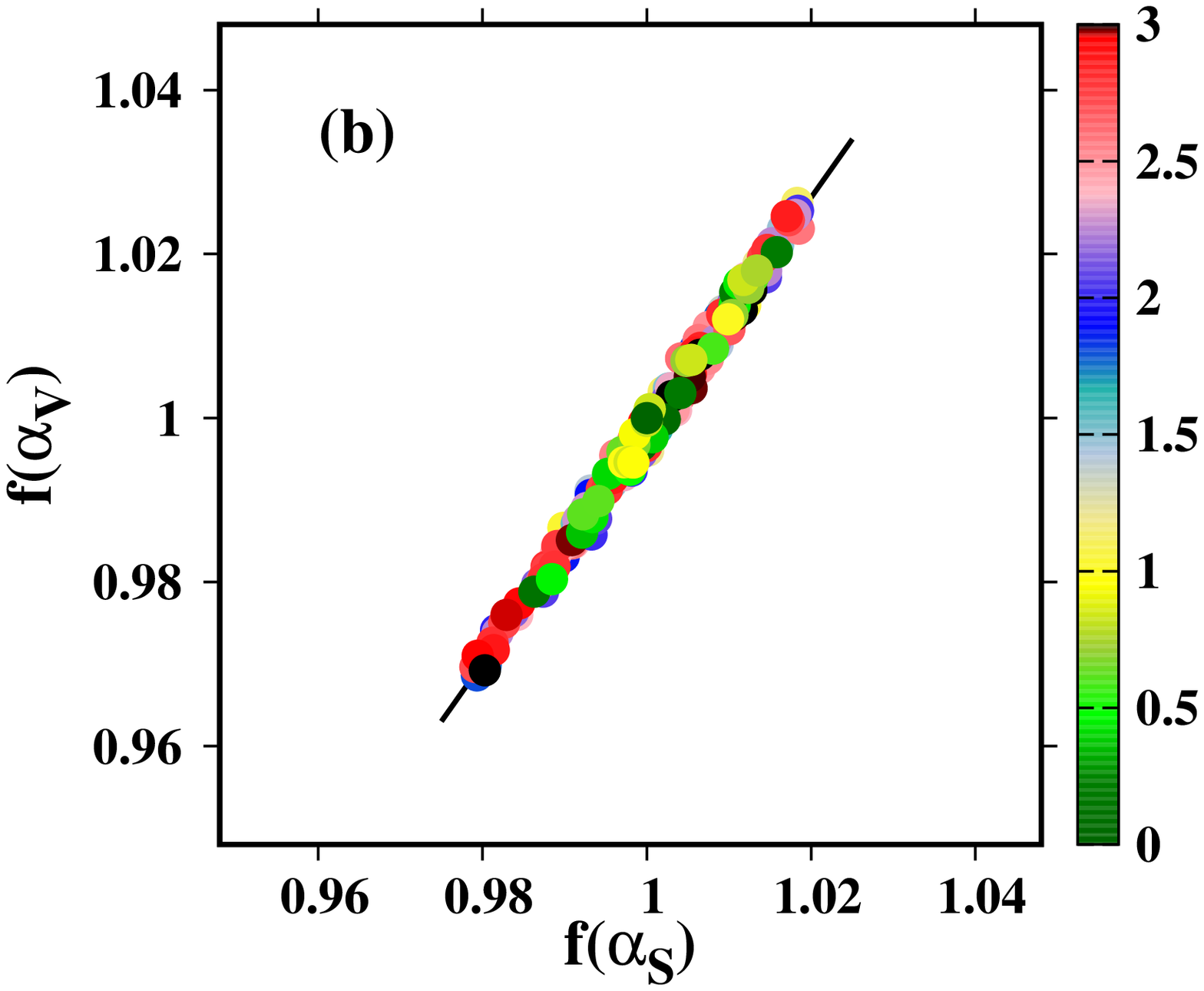}
\includegraphics[angle=-90,width=4.0cm]{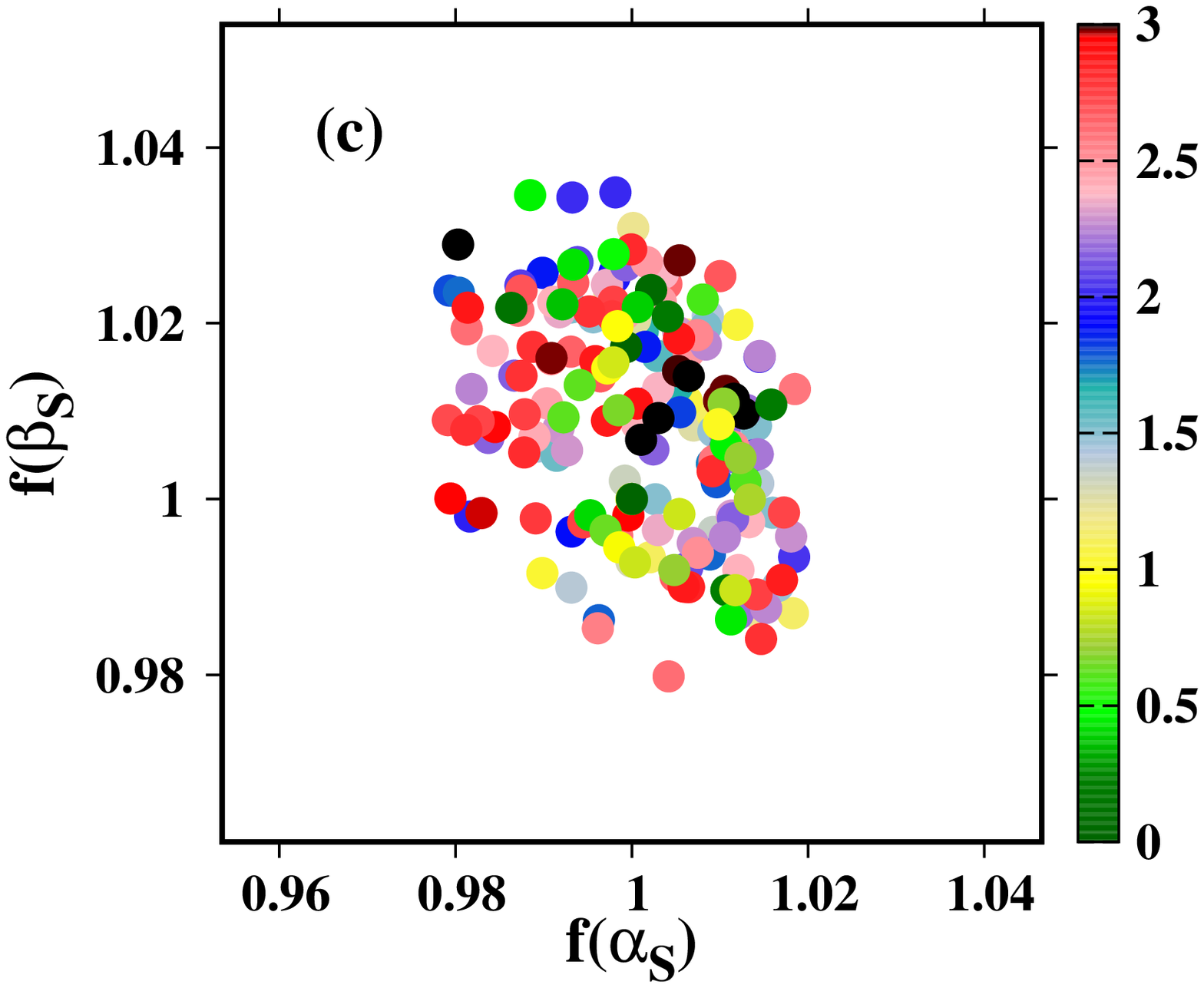}
\includegraphics[angle=-90,width=4.0cm]{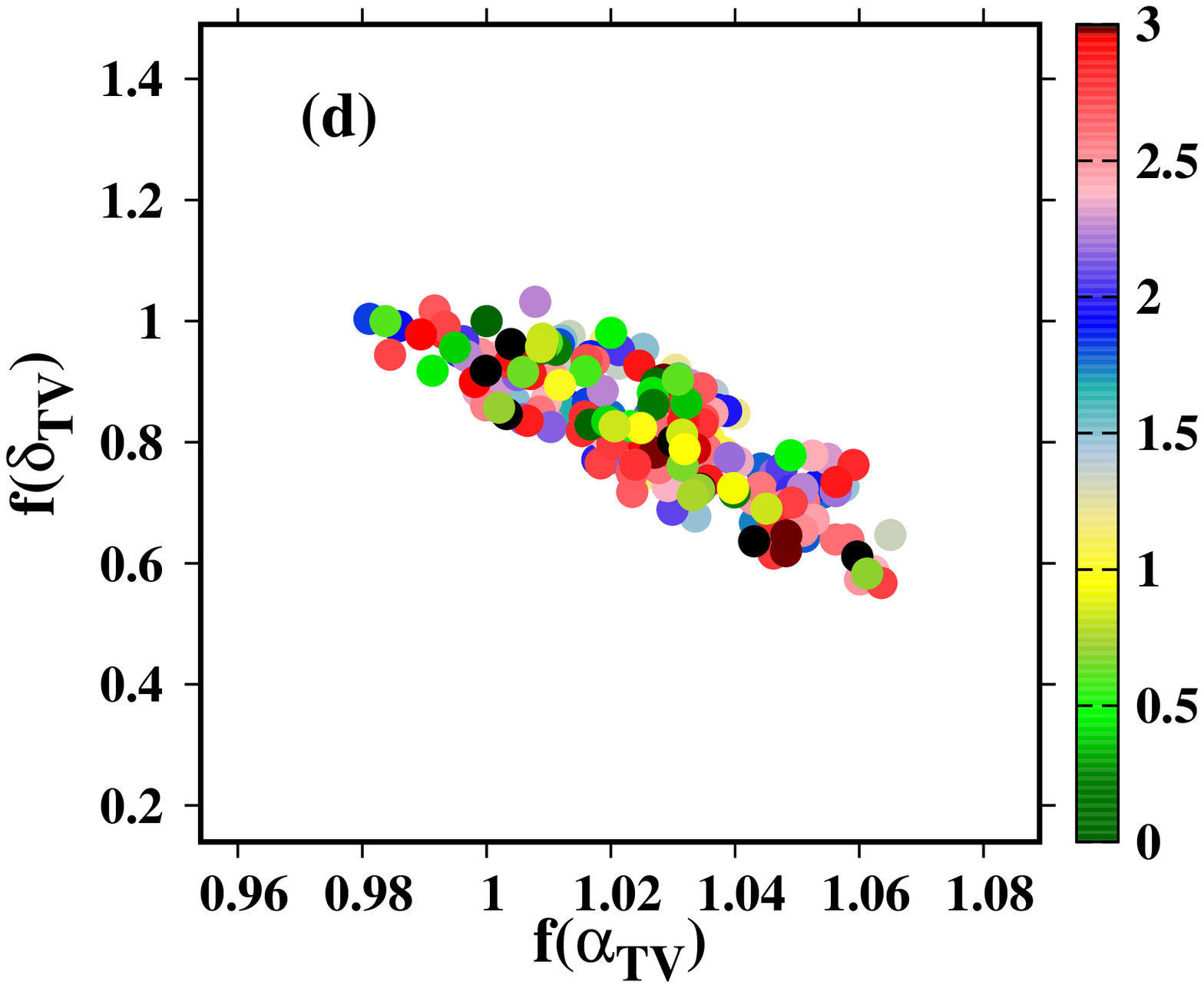}
\includegraphics[angle=-90,width=4.0cm]{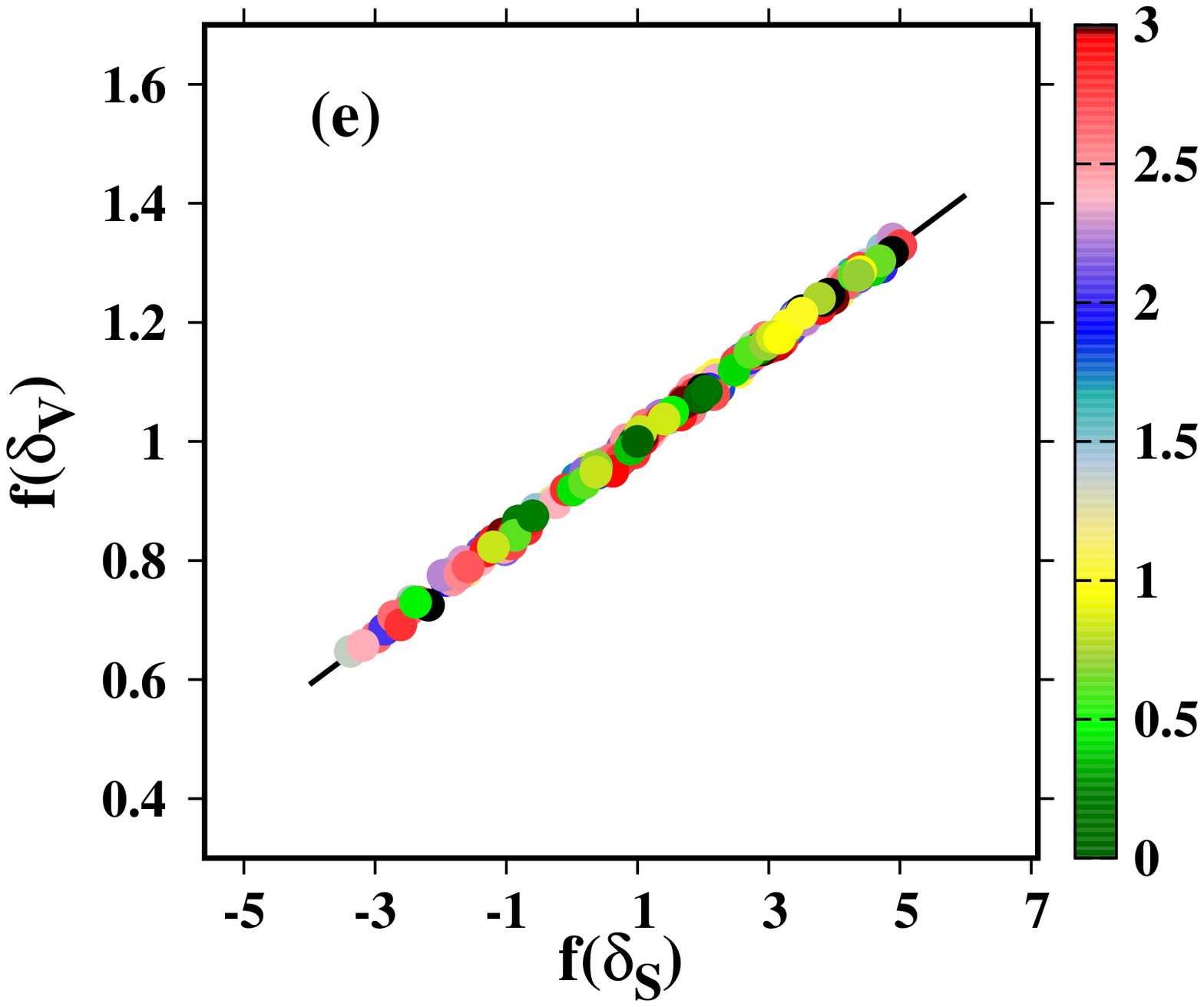}
\includegraphics[angle=-90,width=4.0cm]{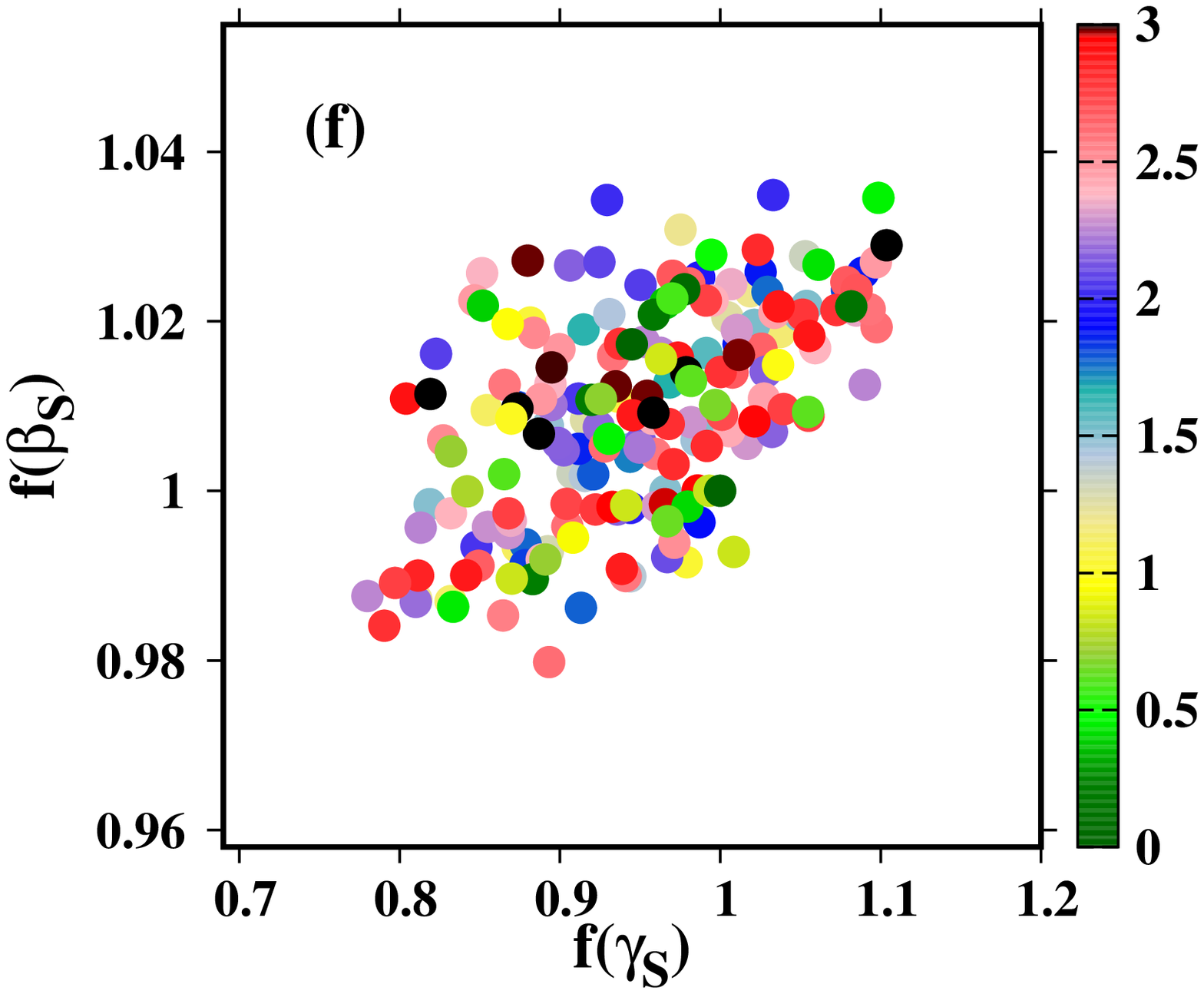}
\includegraphics[angle=-90,width=4.0cm]{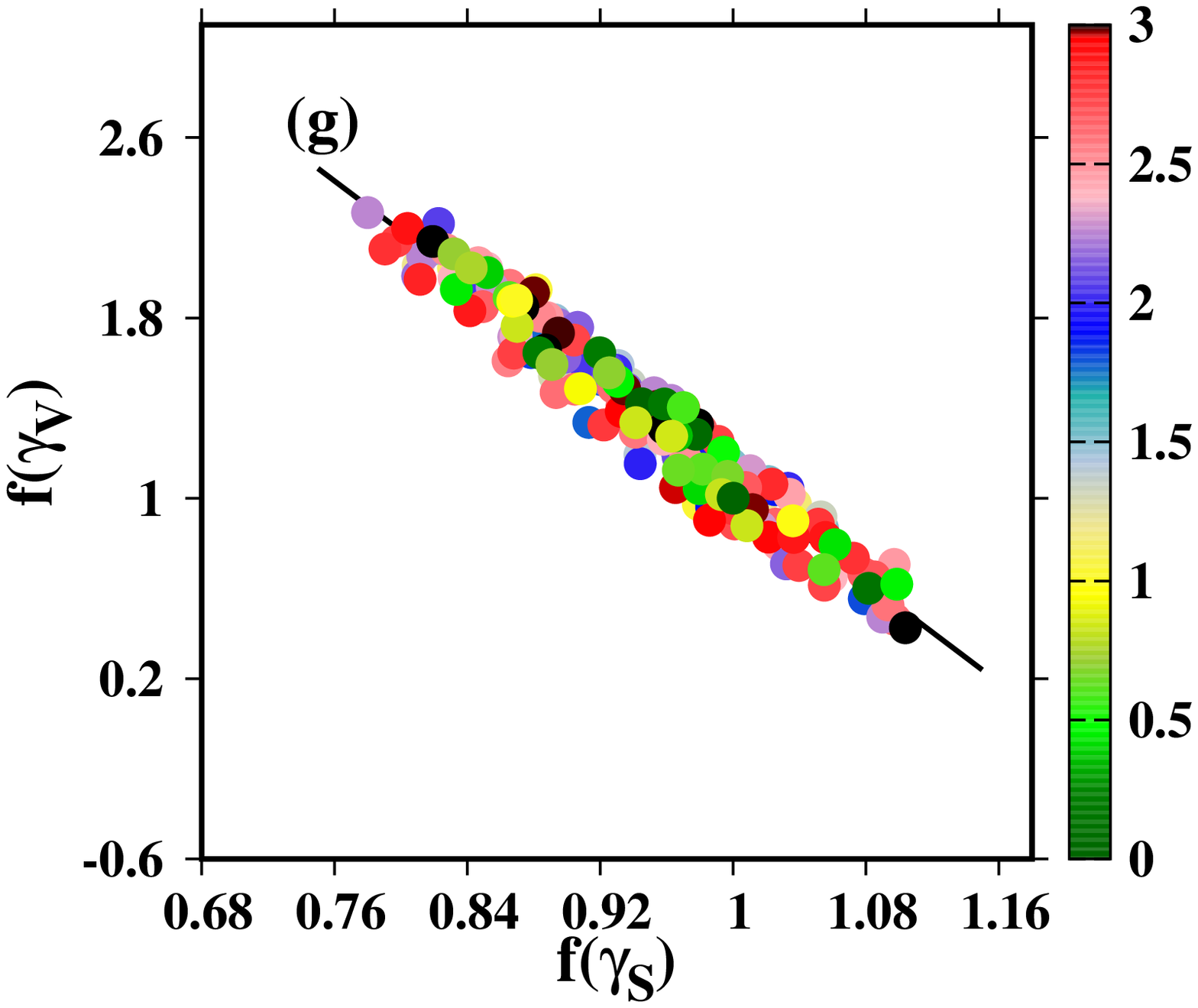}
\includegraphics[angle=-90,width=4.0cm]{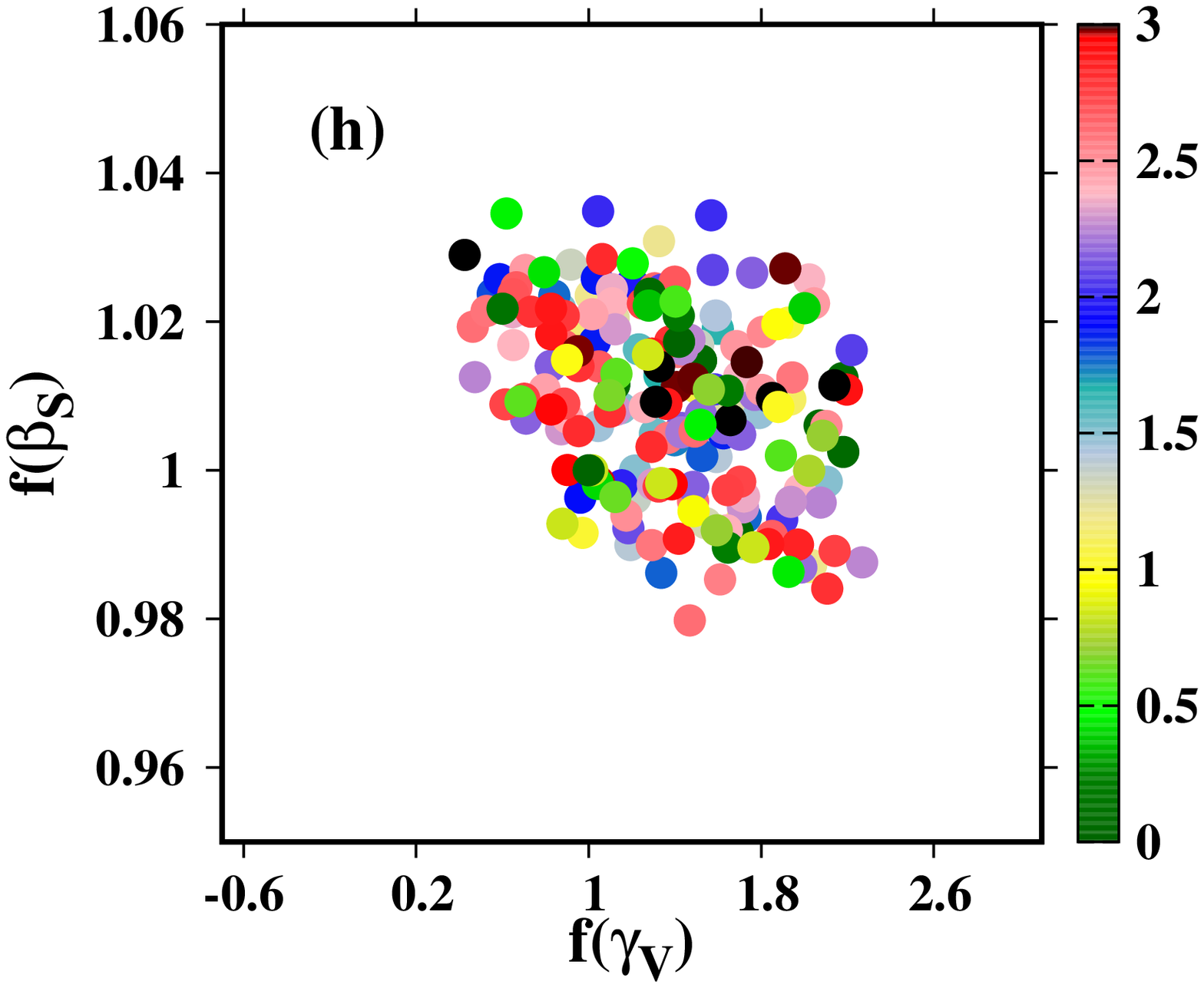}
\caption{The same as Fig.\ \ref{fig-DDME-X-stat} but for the functional PC-X.
}
\label{fig-PC-X-stat}
\end{figure*}

  The functional variations under consideration are defined by the condition
\begin{eqnarray}
\chi^2_{norm} ({\bf p}) \leq \chi^2_{norm}({\bf p}_0) + \Delta \chi^2_{max} \label{cond}
\label{chi-condit}
\end{eqnarray}
The condition $\Delta \chi^2_{max} = 1.0$ specifies the 'physically reasonable' domain
around  ${\bf p}_0$ in which the parametrization ${\bf p}$ provides a reasonable fit and
thus can be considered as acceptable \cite{Stat-an,DNR.14}.  This condition also allows one
to define statistical errors for the physical observables of interest  (see Refs.\
\cite{DNR.14,AAT.19}).  For example, in the CDFT framework, this was done for the
NL5(*) functionals in Ref.\ \cite{AAT.19}.

 The  NL5(*) functionals contain only 6 parameters and thus the volume of the hyperspace is rather
modest. On the contrary, the DDME-X and PC-X functionals contain 8 and 9 parameters, respectively. This
leads to a drastic increase of the volume of the hyperspace which makes numerical calculations with $\Delta \chi^2_{max} = 1.0$  impossible. Thus, in the present investigation we do not consider statistical errors but rather focus on parametric correlations. As shown in Ref.\ \cite{AAT.19} these correlations between the model parameters are visible even for higher values of $\Delta \chi^2_{max} $. Thus, we
use  $\Delta \chi^2_{max} = 3.0$ for the DDME-X and PC-X functionals.
\begin{figure*}[htb]
\centering
\includegraphics[angle=-90,width=5.9cm]{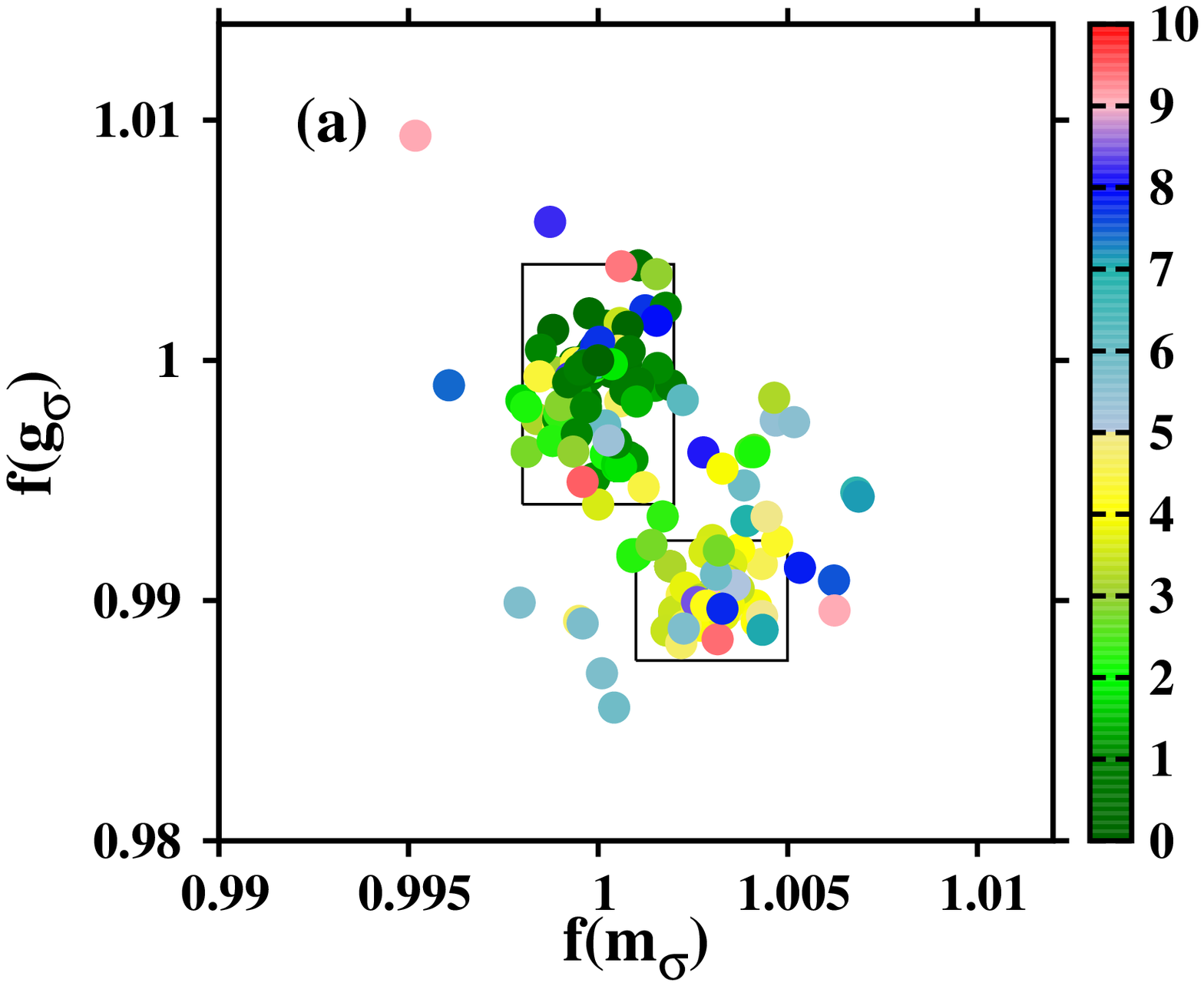}
\includegraphics[angle=-90,width=5.9cm]{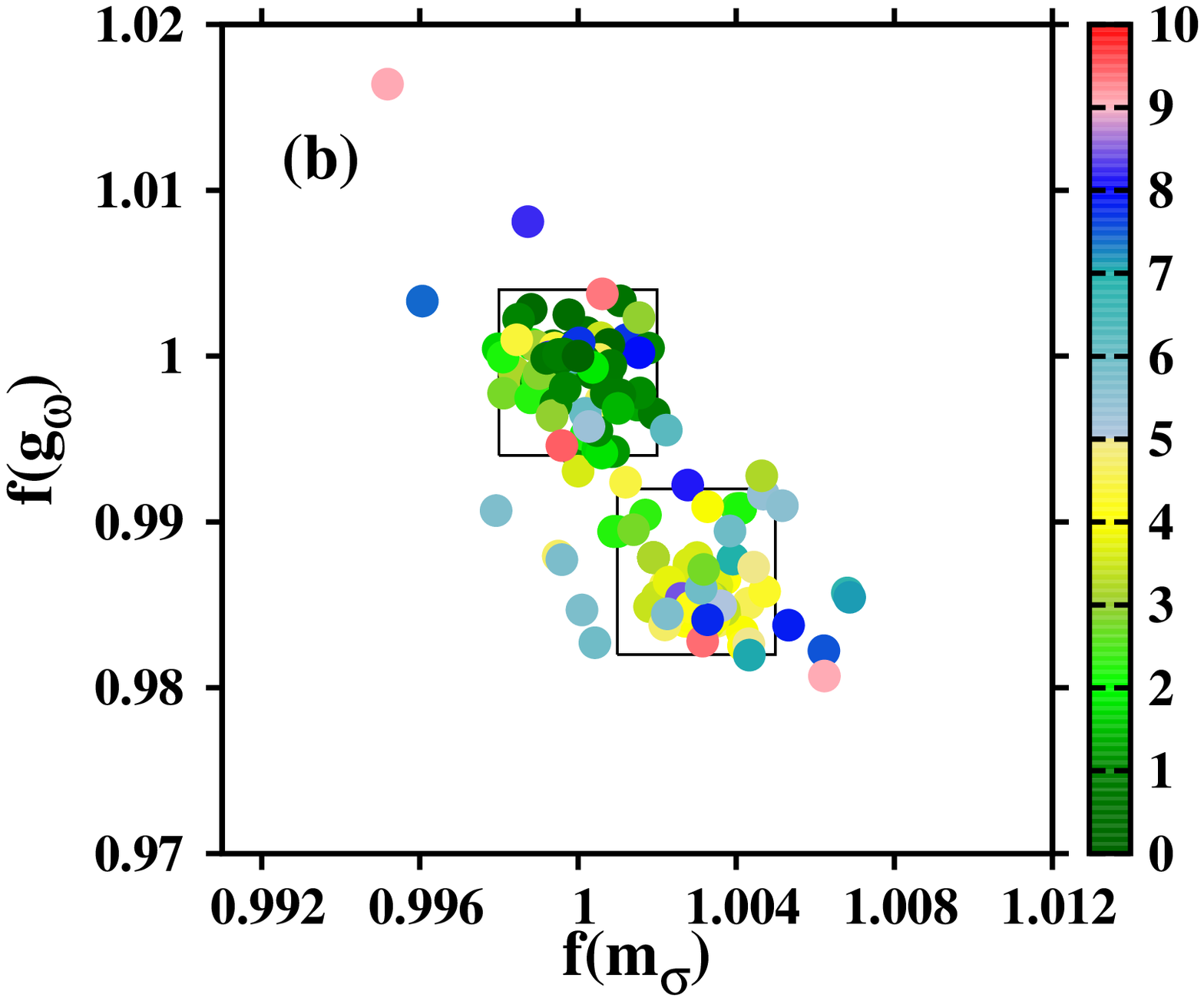}
\includegraphics[angle=-90,width=5.9cm]{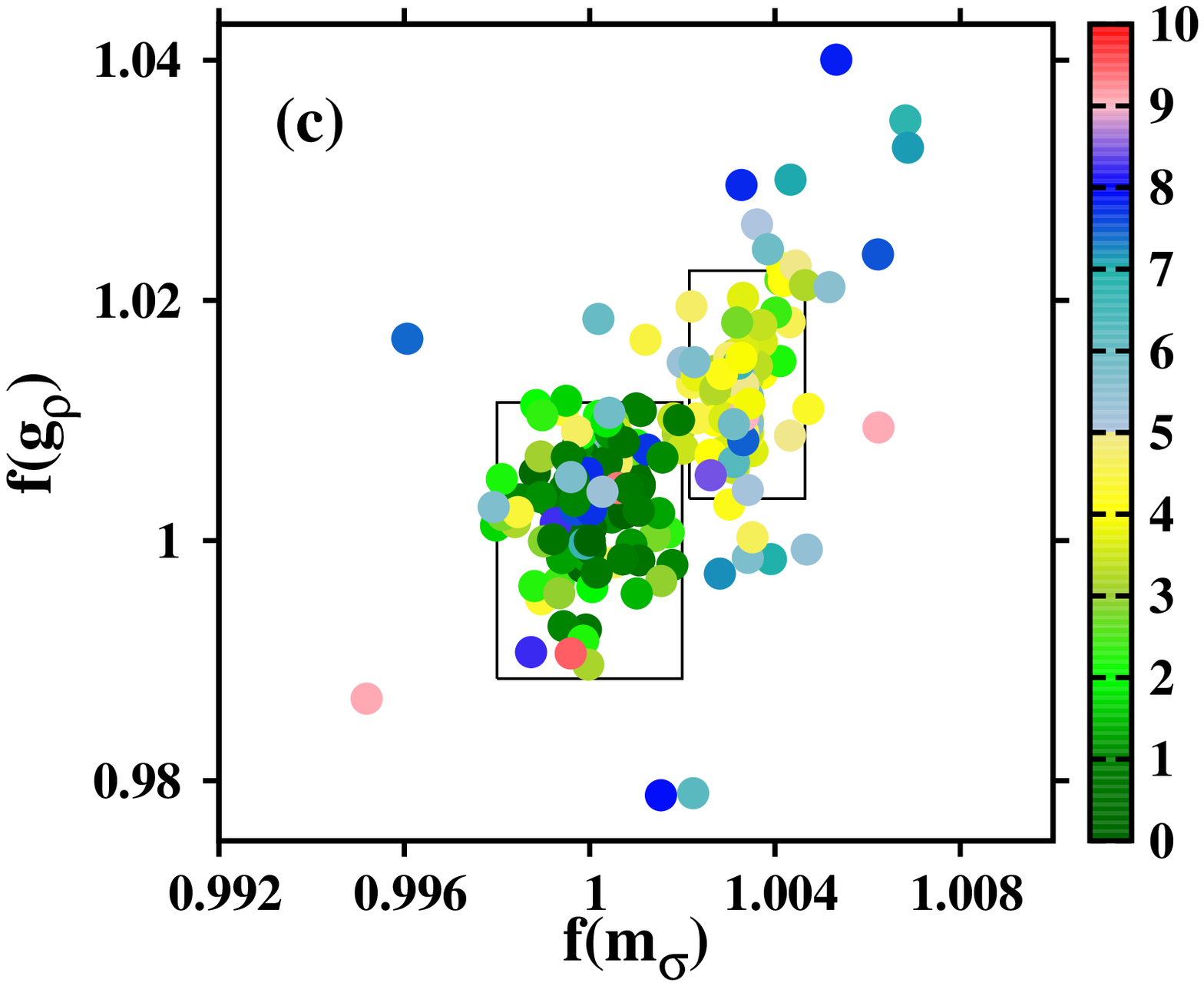}
\includegraphics[angle=-90,width=5.9cm]{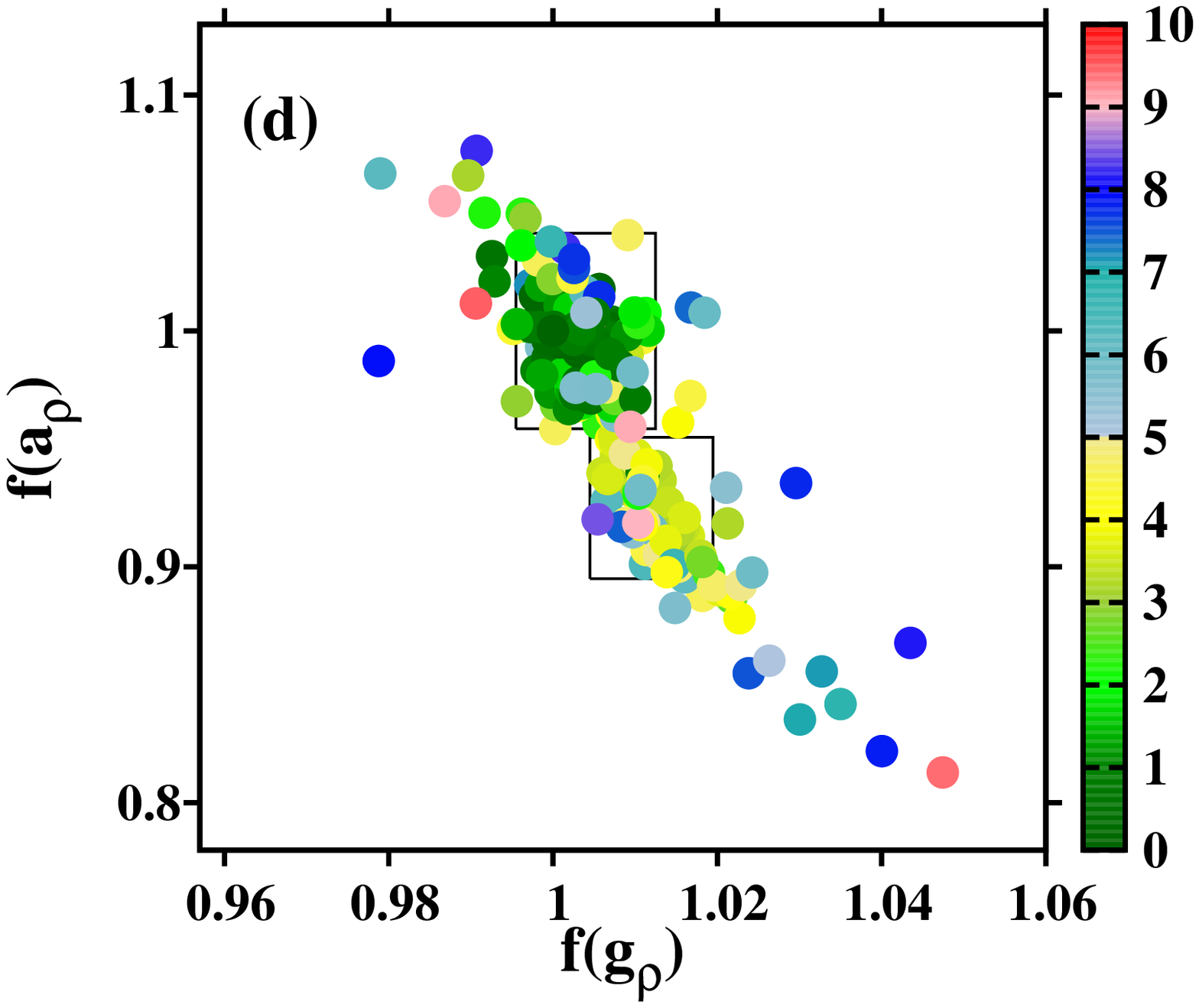}
\includegraphics[angle=-90,width=5.9cm]{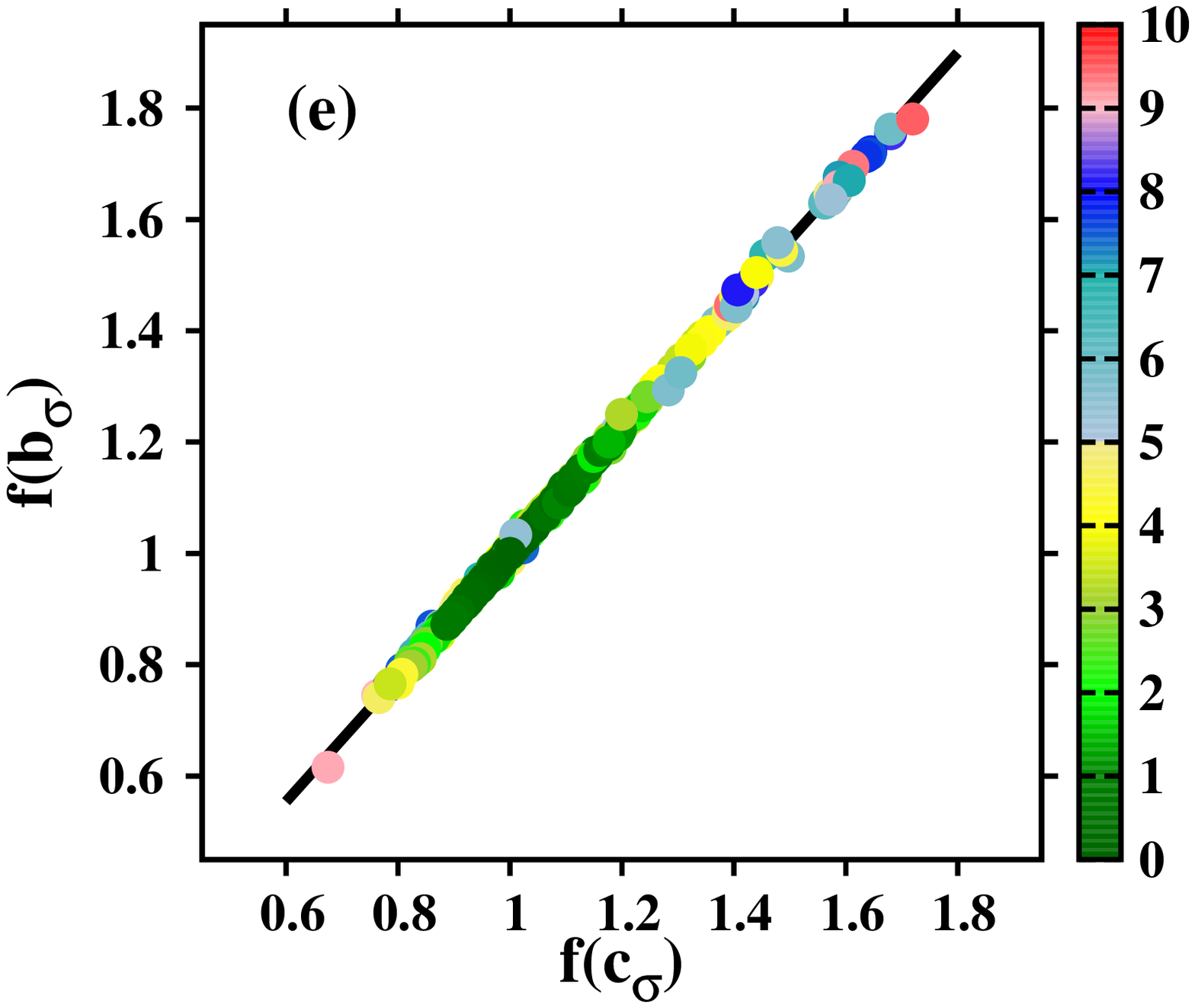}
\includegraphics[angle=-90,width=5.9cm]{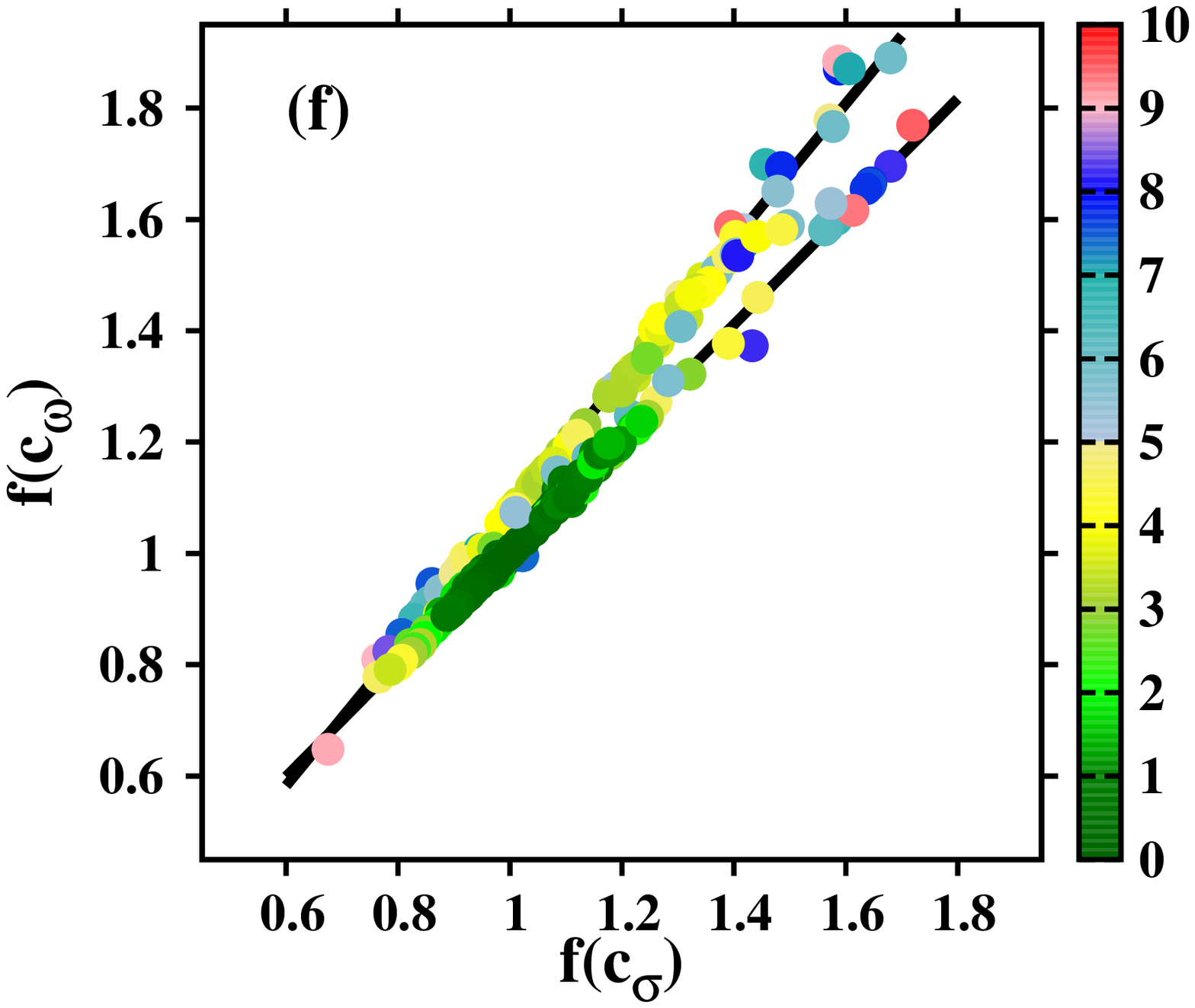}
\caption{Two-dimensional projections of the distribution of the parameters corresponding to
local minima obtained by simplex-based minimizations for the functional DDME-X.
The colors indicate the $\Delta \chi^2$ value of the $\chi^2_{norm}({\bf p})$
for the functionals in these local minima where the latter is expressed as
$\chi^2_{norm}({\bf p}) = \chi^2_{norm}({\bf p}_0) + \Delta \chi^2$. Only local minima
with $\Delta \chi^2 <10.0$ are used here. There are 200 such minima.  The optimal
functional corresponding to the global minimum is located at the intersection of the lines
$f(p_k)=1.0$ and $f(p_j)=1.0$.
}
\label{simplex-DDME-X}
\end{figure*}

\begin{figure*}[htb]
\centering
\includegraphics[angle=-90,width=4.0cm]{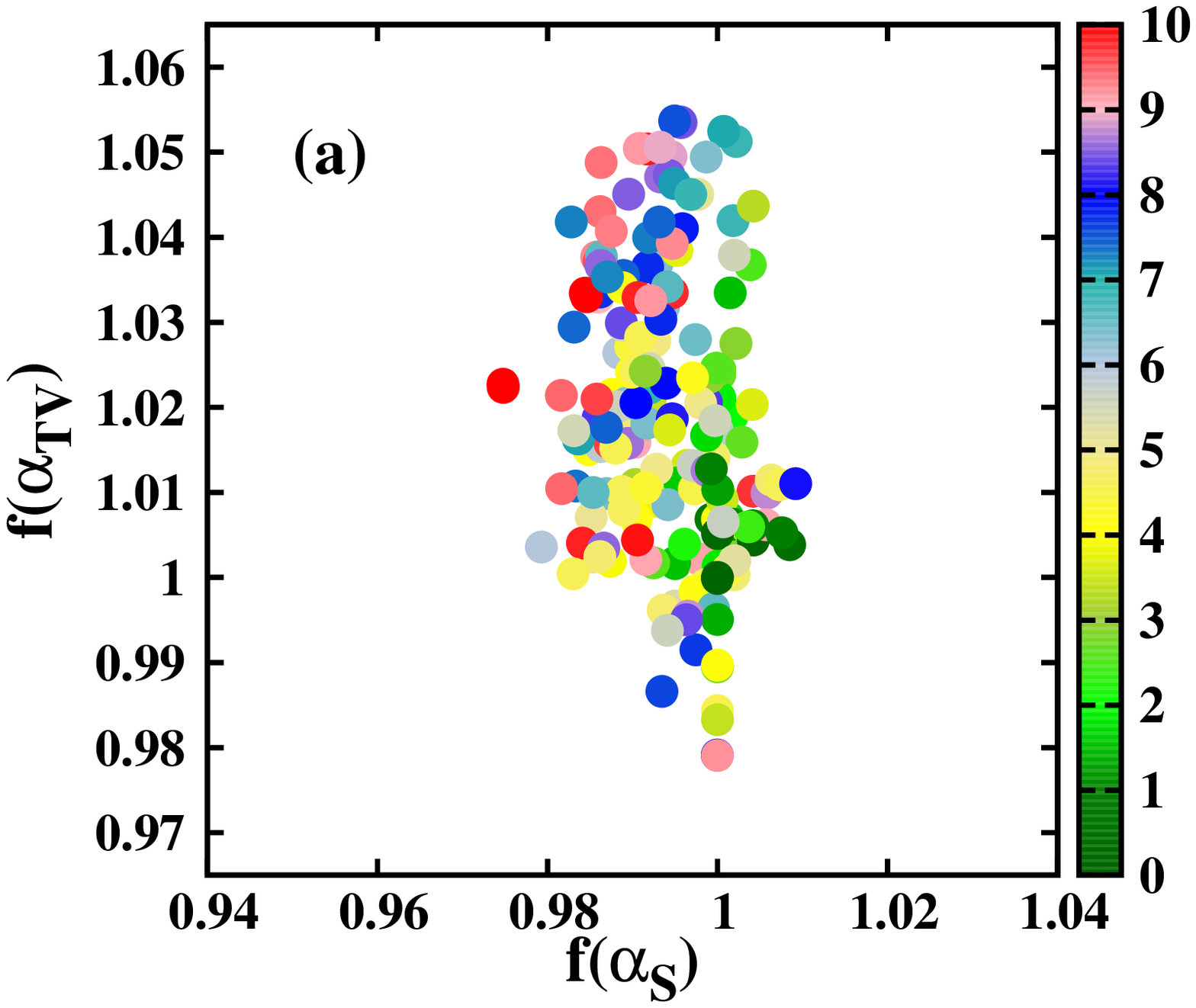}
\includegraphics[angle=-90,width=4.0cm]{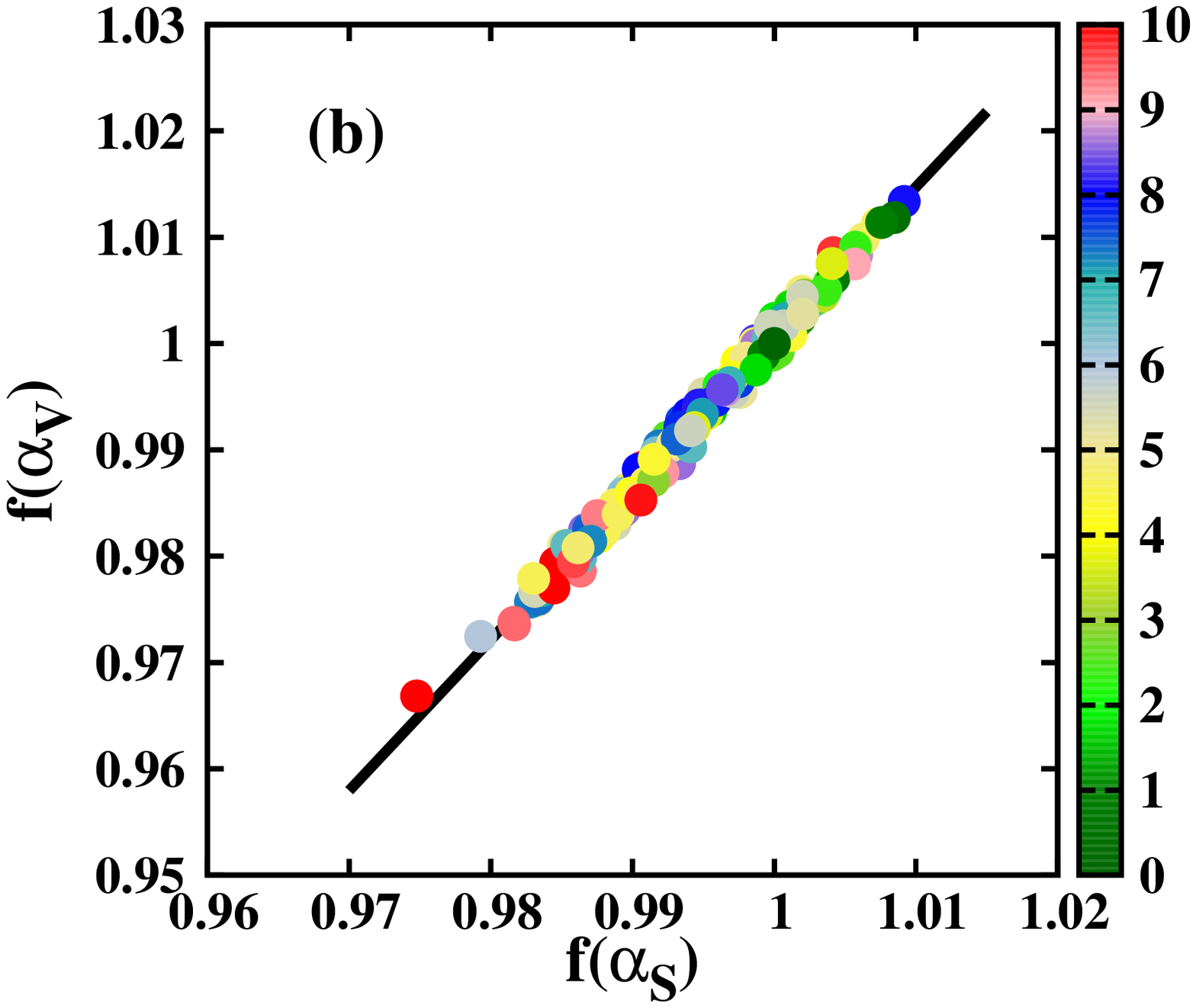}
\includegraphics[angle=-90,width=4.0cm]{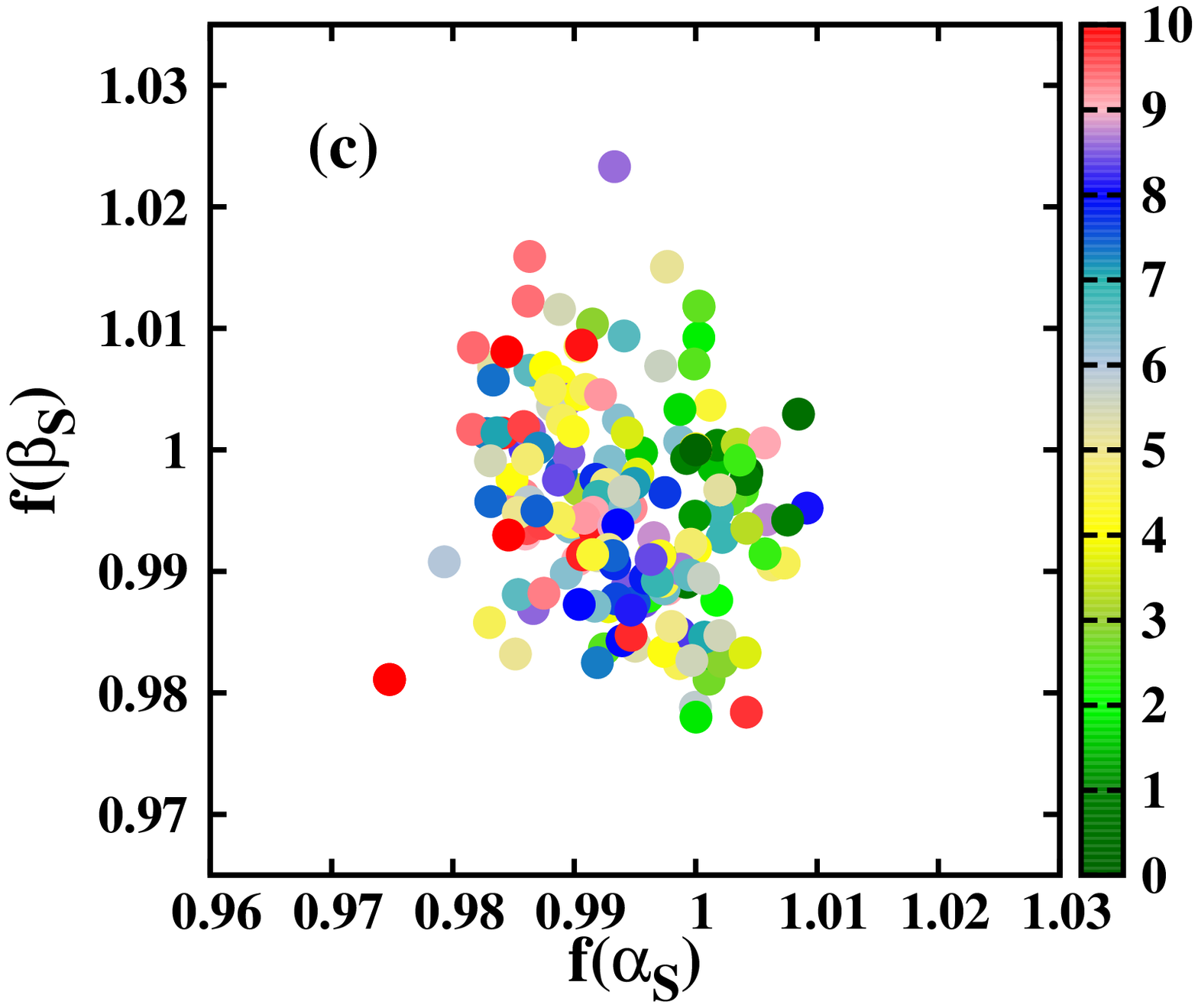}
\includegraphics[angle=-90,width=4.0cm]{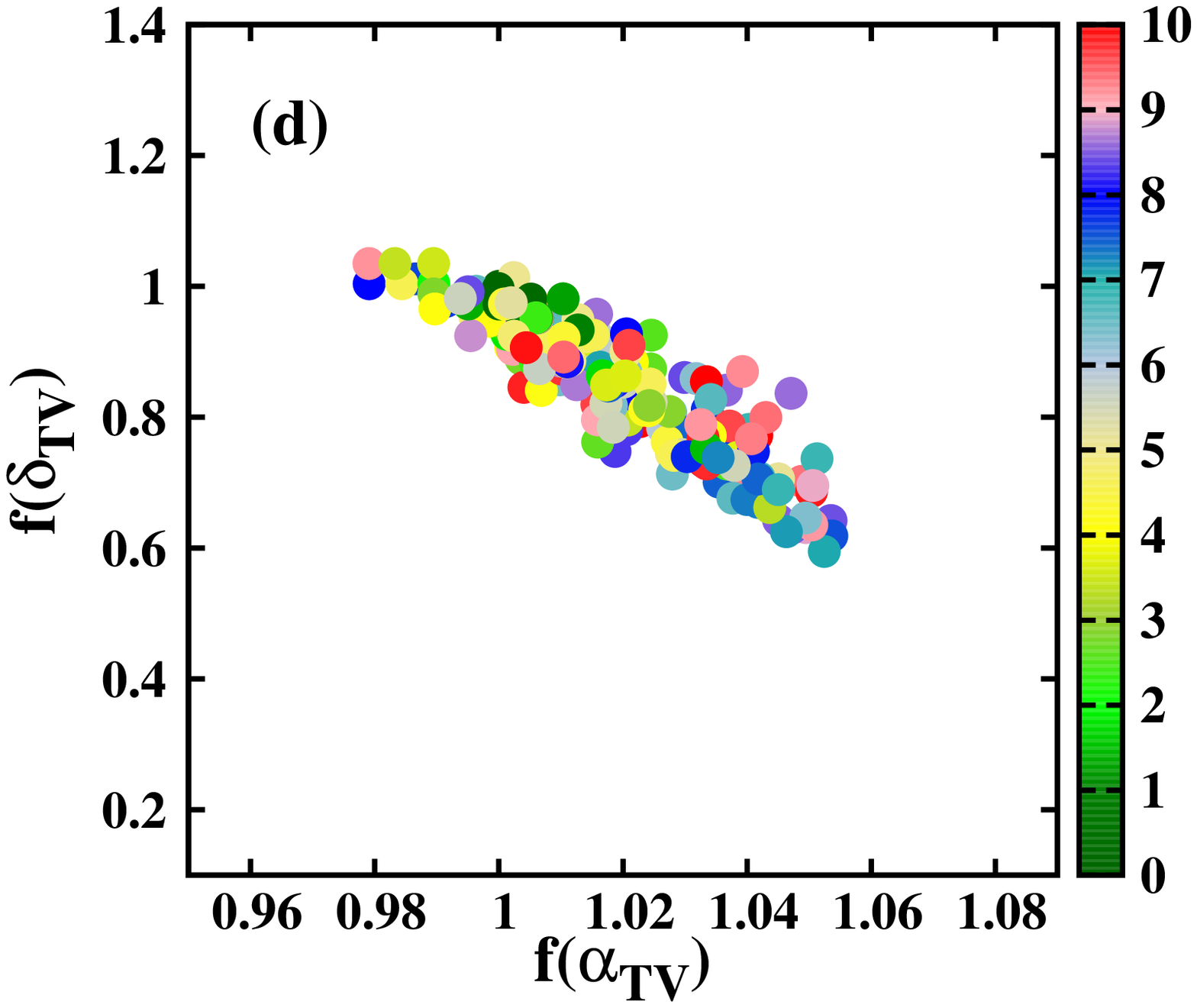}
\includegraphics[angle=-90,width=4.0cm]{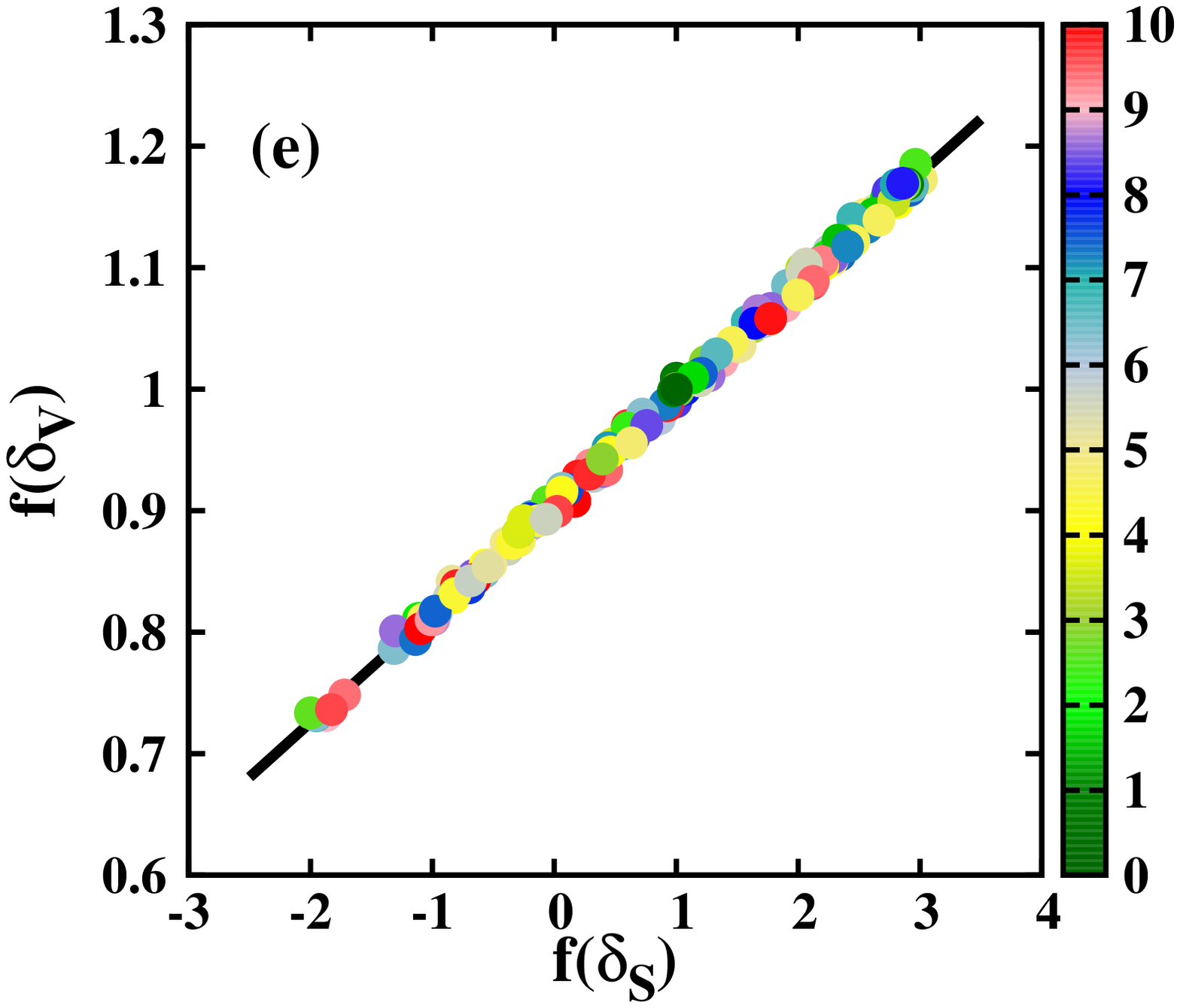}
\includegraphics[angle=-90,width=4.0cm]{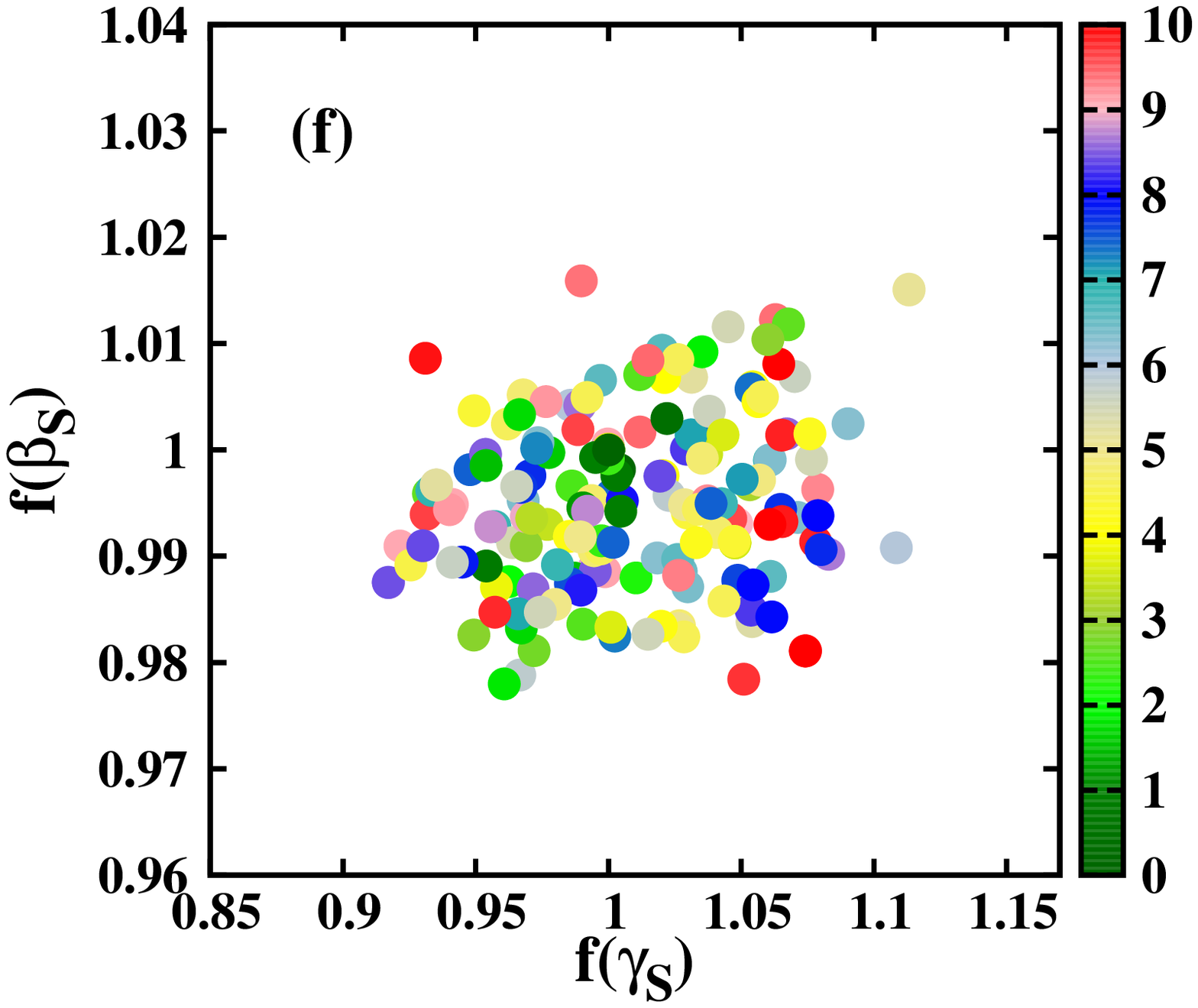}
\includegraphics[angle=-90,width=4.0cm]{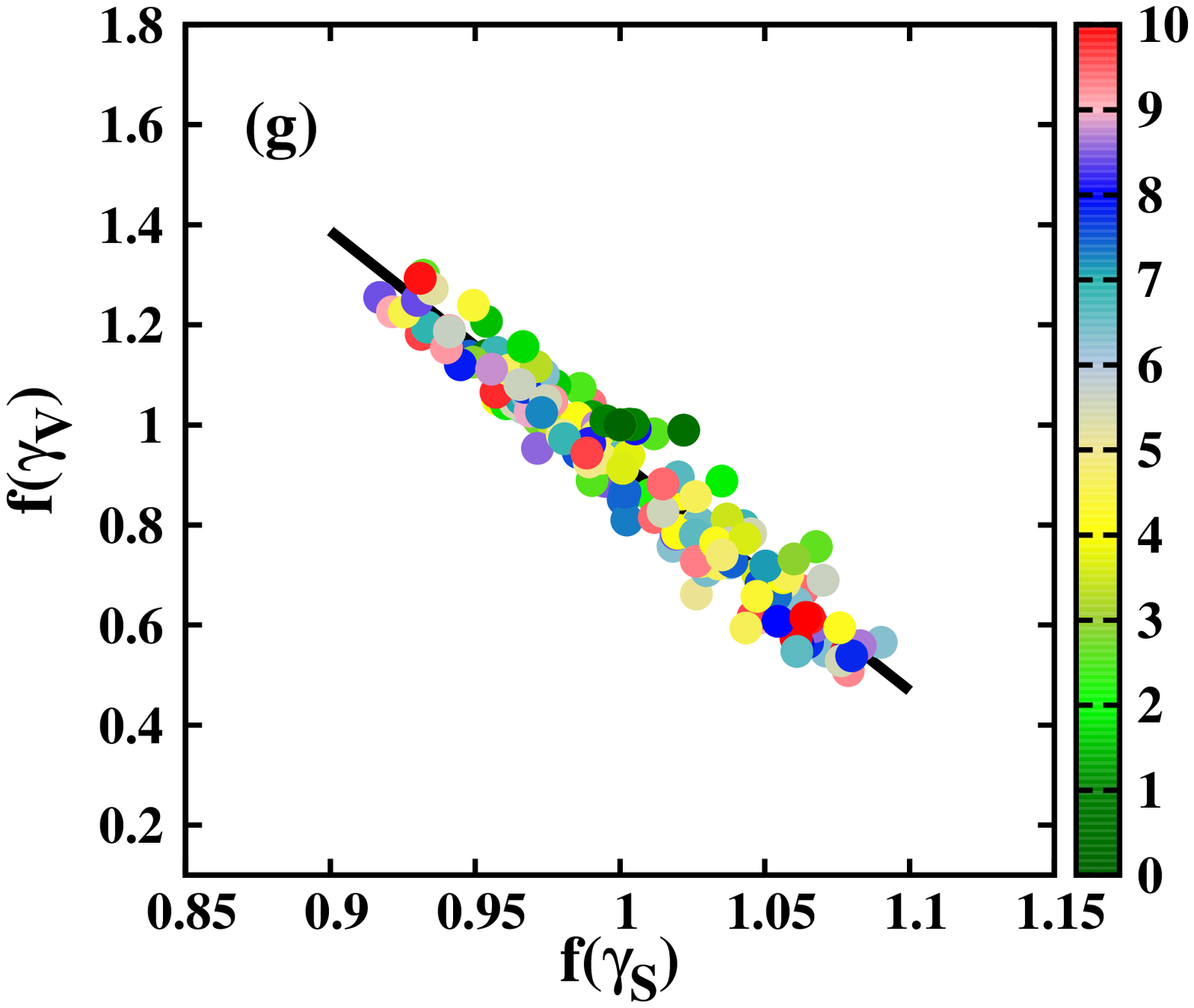}
\includegraphics[angle=-90,width=4.0cm]{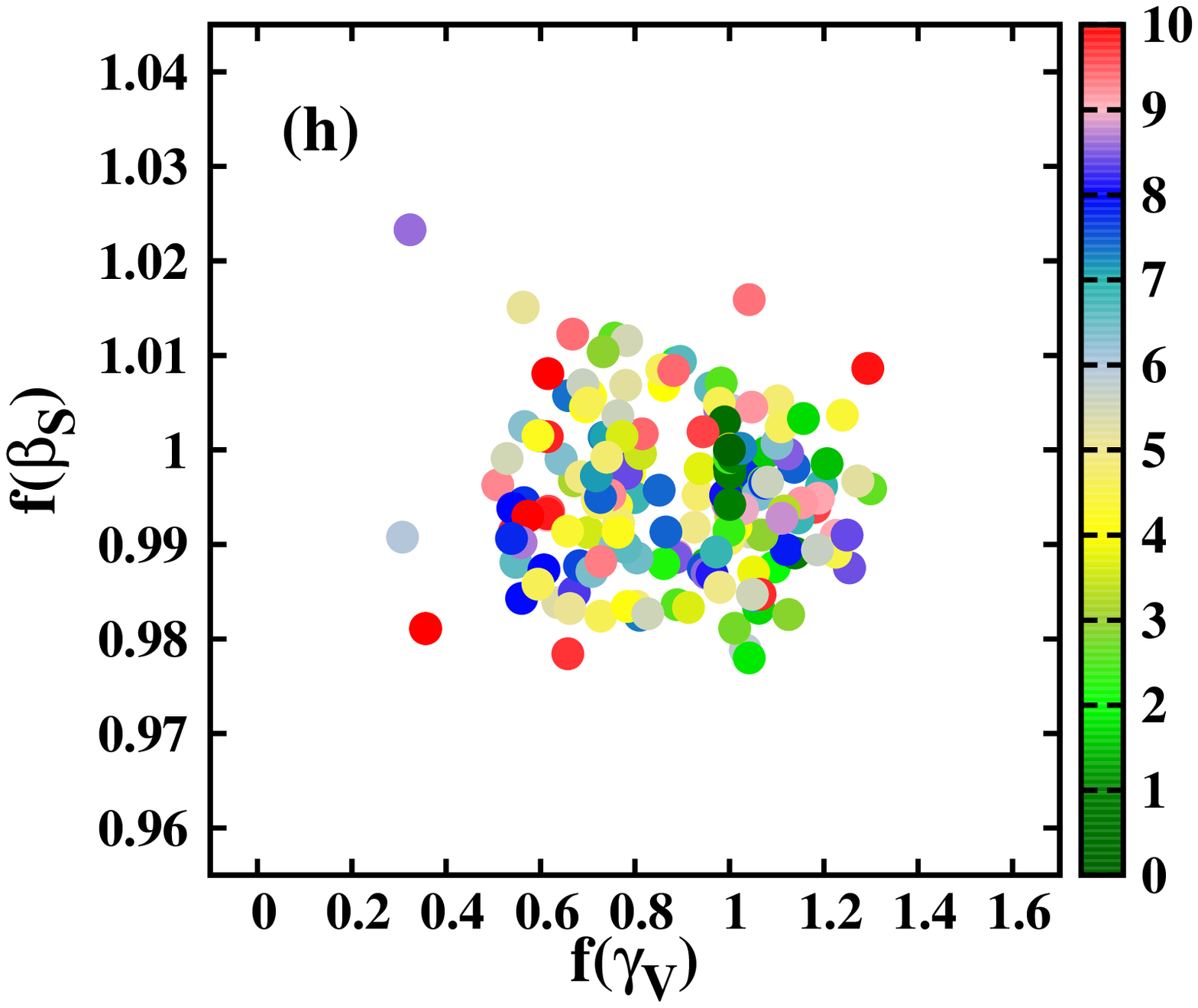}
\caption{The same as Fig.\ \ref{simplex-DDME-X} but for the PC-X functional.
\label{simplex-PC-X}
}
\end{figure*}

The numerical calculations are performed in the following way: New parametrizations ${\bf p}=(p_k,k=1,N)$ are randomly generated in the $N=N_{par}$-dimensional parameter hyperspace  and they are accepted if the condition  (\ref{cond}) is satisfied. The domain in the $N=N_{par}$-dimensional parameter hyperspace, in which the calculations
are performed, is defined as
${\bf P_{space}}=[p_{1_{min}} - p_{1_{max}}, p_{2_{min}} - p_{2_{max}}, ..., p_{N_{min}} - p_{N_{max}}]$, where $p_{k_{min}}$ and $p_{k_{max}}$ represent the lower and upper boundaries for the
variation of the $k-th$ parameter.  These boundaries are defined in such a way that their further
increase (for $p_{k_{max}}$) or decrease (for $p_{k_{min}}$) does not lead to additional points in
parameter hyperspace which satisfy Eq.\ (\ref{cond}).

  Note that, in the following, instead of the functional parameters $p_k$ ($k=1,N$) we are
  using the ratios (see Ref.\ \cite{AAT.19})
\begin{equation}
f(p_k)=\frac{p_k}{p_k^{opt}}
\end{equation}
where $p_k^{opt}$ is the value of the parameter in the optimal functional and
$k$ indicates the type of the parameter. This allows one to understand the range
of the variations of the parameters and related parametric correlations in the
functionals.

   In Fig.\ \ref{fig-DDME-X-stat} we consider the CEDF DDME-X and show, for the
randomly generated parameters obeying the condition (\ref{chi-condit}), the
2-dimensional distributions of indicated pairs of the parameters.
The parameters vary with respect to the central value of the distribution (which are typically given
by the parameters of the optimal functional) by at most 0.5\% for $m_\sigma$,
0.6\% for $g_\sigma$, 1\% for $g_{\omega}$, 2.5\% for $g_{\rho}$, 10\%
for a$_{\rho}$, and 30\% for $c_{\sigma}$, $b_\sigma$ and $c_{\omega}$.
Similar plots are presented in Fig.\ \ref{fig-PC-X-stat} for the PC-X functional.

  One can speak of parametric correlations between these parameters when
one parameter $p_k$ can, with a reasonable degree of accuracy, be expressed
as a function of other parameters, for example, as a
function of the parameter $p_j$. The simplest type of the
correlations is a linear one as given by
\begin{equation}
f(p_k) = a f(p_j)  + b
\label{param-correl}
\end{equation}
For example, the following linear relations exist between the parameters
of the  DDME-X functional (shown by solid black lines in Figs.\ \ref{fig-DDME-X-stat}e and \ref{fig-DDME-X-stat}f)
\begin{eqnarray}
f(b_{\sigma}) &=& 1.1396 f(c_{\sigma})  - 0.14191 \nonumber\\
f(c_{\omega}) &=& 1.083 f(c_{\sigma})  - 0.08655
\label{eq-stat-DDME-X}
\end{eqnarray}
and between the parameters of the PC-X functional
(shown by solid black lines in Figs.~\ref{fig-PC-X-stat}b, \ref{fig-PC-X-stat}e, and \ref{fig-PC-X-stat}g)
\begin{eqnarray}
f(\alpha_{v}) &=& 1.4203 f(\alpha_{s})  -  0.42178 \nonumber \\
f(\delta_{v}) &=& 0.08221 f(\delta_{s}) + 0.96062 \nonumber \\
f(\gamma_{v}) &=& -5.5582 f(\gamma_{s}) + 6.6311
\label{eq-stat-PC-X}
\end{eqnarray}
In the case of non-linear functionals  linear relations exits between
$g_2$ and $g_3$ which define, in Eq. (\ref{lagrnl5}), the density dependence
of the functional (see Ref. \cite{AAT.19}).

  Because of the two linear correlations (\ref{eq-stat-DDME-X}) for the functionals DDME-X
and because of the three linear correlations (\ref{eq-stat-PC-X}) for the functionals PC-X
the number of independent parameters can be reduced from 8 to 6 in the functional DDME-X
and from 9 to 6 in the functional PC-X.
Note that the accounting of the parametric correlations in the case of non-linear meson-exchange
models leaves only 5 independent parameters (see Ref.\ \cite{AAT.19}). Thus, one can conclude that the ground state
and nuclear matter properties usually used in the fitting protocols allow one to define only 5-6
(dependent on the model structure) independent parameters in the case of CDFT. Models
with a larger number of parameters are most likely over-parametrized.

   These results are consistent for the three models. For the NLME model
we have only a density dependence in the isoscalar channel. Originally
it is determined by 2 parameters $g_2$ and $g_3$. The parametric correlations
lead to a reduction to only one parameter for the density dependence in
the isoscalar channel. The density dependence in the isovector channel is
neglected and this obviously leads to unphysically large values of
the slope of the symmetry energy $L_0$ (see Ref.\ \cite{AA.16}).
In the DDME model, we have originally 3 parameters in the isoscalar channel
and one parameter in the isovector channel. We found no parametric correlations
in the isovector channel, but the number of parameters in the isoscalar channel
is reduced by parametric correlations from 3 to 1. In the PC-models we have
also one parameter in the isovector channel, but the number of parameters in
the isoscalar channel is reduced from 4 to 1. Finally we have in all cases
one parameter in the isoscalar channel and one parameter in the isovector
channel.

  This result can be understood qualitatively also on a microscopic basis.
Starting from the bare nucleon-nucleon interaction adjusted to the
nucleon-nucleon scattering data~\cite{Machleidt1987_PR149-1} and using relativistic Brueckner-Hartree-Fock theory
in symmetric and asymmetric nuclear matter at various densities one is able to
derive the relativistic self-energies of nucleons in nuclear matter without any
phenomenological parameters
\cite{Marcos1989_PRC39-1134,Sehn1990_NPA519-289,BT.92,
Haddad1993_PRC48-2740,Fritz1994_PRC49-633}.
By adjusting the self-energies obtained from CDFT in nuclear matter at the same density
one is able to derive the density dependence of the coupling constants in a microscopic
way~\cite{BT.92}. However, in the Brueckner calculations, a number of approximations have been used
and therefore this mapping is not unique. At present, the results obtained from such calculations in
finite nuclei are rather different and, so far, their quality is far from that obtained with
phenomenological CDFTs (see, for instance, Fig. 11 in Ref. \cite{SHEN-SH2019}). However, they
all show in the isoscalar channel a density dependence in the relevant density interval
between 0.5 and 1.1 of the saturation
density, which is close to a linear density dependence (see, for instance, Refs. \cite{terHaar1987_PR149-207,Brockmann1990_PRC42-1965,Serra2005_PTP113-1009}).
This fact gives at least a qualitative explanation, why the parametric correlations
discussed here allow a reduction to one parameter in the isoscalar channel.

   In the isovector channel, there is no reduction of the number of parameters describing the 
density dependence, because, from the beginning, we have 
no density dependence in the PC-X CEDF and in the non-linear meson
coupling models (such as the NL5 family of CEDFs) and only one parameter for the 
density dependence in the DDME-X functional. 
This is easy to understand because the effects in the isovector channel are much smaller than those 
in the isoscalar channel in which two huge scalar and vector fields $S$ and $V$
cancel in the nucleonic potential and add up in the spin-orbit one. 
As it has been shown in Ref. \cite{DD-MEdelta}, present data for ground states of finite 
nuclei do not allow to distinguish corresponding scalar and vector potentials in the isovector 
channel.  

It is necessary to recognize that the search for parametric correlations in the
multidimensional parameter hyperspace by the method described above is extremely
time-consuming even with modern high performance computers. Thus, we looked for
alternative methods for such a search.  The simplest method we found is based on
the minimization by the simplex method (see Ref.\ \cite{NumRec}). However, minimizations
by the simplex method are prone to stack in local minima and that is a reason why it is
not recommended for the search for global minimum. However, in the context of the search
of parametric correlations the drawback becomes an advantage. Starting from different randomly
defined parameter vectors we perform a number of trial minimizations with the simplex method.
They lead to different local minima in the parameter hyperspace. The distributions of the
parameters corresponding to these local minima are shown in Figs.\ \ref{simplex-DDME-X}
and \ref{simplex-PC-X} for the functionals DDME-X and PC-X, respectively.
One can see that the parametric correlations seen in Figs.\ \ref{fig-DDME-X-stat}
and \ref{fig-PC-X-stat} are also clearly visible in these two figures. It is important to note that the
search of parametric correlations via the simplex-based minimization method
is at least by an order of magnitude less time-consuming than a fully statistical
search based on Eq.\ (\ref{chi-condit}) as it is shown in Figs. \ref{fig-DDME-X-stat} and \ref{fig-PC-X-stat}.

  It is also important that the simplex-based minimization method allows one to find a fine
structure of such correlations which can be hidden in a fully statistical approach. This
is illustrated in Fig.\ \ref{simplex-DDME-X}. Figs.\ \ref{simplex-DDME-X}a-d show the
coexistence of two long-range structures corresponding to a global and a sub-global
minima;  the respective parameter ranges are enclosed by the rectangles in panels
(a-d). While the parametric correlations between the parameters $b_{\sigma}$ and $c_{\sigma}$
are the same in both structures [which is not surprising considering that these
two parameters describe the same type of meson] (see Fig.\ \ref{simplex-DDME-X}e),
they are different between the $c_{\omega}$ and $c_{\sigma}$ parameters for these
long-range structures (see Fig.\ \ref{simplex-DDME-X}e).  This is also a reason why
the correlations between the latter two parameters are broader (in width) in the fully statistical
analysis presented in Fig.\  \ref{fig-DDME-X-stat}e; this is because
$\Delta \chi^2_{max} = 3.0$ used in this analysis covers both long-range structures.

   The linear correlations (shown by black lines in Figs. 3 and 4) defined via
the simplex-based minimization method are given by
\begin{eqnarray}
f(b_{\sigma}) &=& 1.1212 f(c_{\sigma})  - 0.11845 \nonumber \\
f(c_{\omega}) &=& 1.0149 f(c_{\sigma})  - 0.01002 \nonumber \\  
f(c_{\omega}) &=& 1.2254 f(c_{\sigma})  - 0.15263  
\label{eq-simpl-DDME-X}
\end{eqnarray}
for the DDME-X functional.   Note that the values given in the second and third lines
of  Eq.\ (\ref{eq-simpl-DDME-X}) correspond to the global and sub-global minima
of the $\chi^2$ function, respectively. The equations
\begin{eqnarray}
f(\alpha_{v}) &=&1.419  f(\alpha_{s}) - 0.41846 \nonumber \\
f(\delta_{v}) &=&0.09026  f(\delta_{s}) + 0.90639 \nonumber \\
f(\gamma_{v}) &=&-4.5936 f(\gamma_{s}) + 5.9217
\label{eq-simpl-PC-X}
\end{eqnarray}
define similar correlations between the parameters of the PC-X functional.
One finds an extreme similarity of the parametric correlations for PC-X
obtained via the simplex-based minimization method (Eq. (\ref{eq-simpl-PC-X}))
and those defined from full statistical analysis (Eq.\ (\ref{eq-stat-PC-X})).  The
same is true for the correlations between the parameters $b_{\sigma}$ and $c_{\sigma}$
of the DDME-X functional (compare the upper lines of Eqs.\
(\ref{eq-stat-DDME-X})  and (\ref{eq-simpl-DDME-X})). However, the results for
the parametric correlations between the parameters $c_{\omega}$  and $c_{\sigma}$
 of the DDME-X functional obtained by full statistical analysis are
located in between those defined by means of the simplex-based minimization
method (compare Eqs.\ (\ref{eq-stat-DDME-X})  and (\ref{eq-simpl-DDME-X})).
This is due to the fact that because of the selection of the $\Delta \chi^2_{max}$
value the results obtained with former method are an "envelope" of those obtained
with latter method.

  In the  context of the analysis of theoretical uncertainties there is one clear
advantage in the reduction of the dimensionality of the parameter hyperspace
via the removal of parametric correlations: such a reduction leads to a
decrease of the statistical errors \cite{DD.17,AAT.19}.

   In conclusion, density functional theories (DFT) are defined by underlying functionals.
Some of those functionals depend on a substantial number of parameters. However,
with the exception of non-linear meson-exchange CEDFs~\cite{AAT.19} the
parametric correlations between them have not been studied before.
Using covariant DFT as an example and statistical tools, we have investigated  
such correlations for major classes of covariant energy density functionals for the first time.
These include the non-linear meson-exchange functionals (NLME) studied in Ref.~\cite{AAT.19}
and the functionals DDME-X and PC-X studied for the first time in the present manuscript.
These functionals are defined by the ground state properties of spherical nuclei and with exception
of PC-X by the pseudodata on nuclear matter. It turns out
that parametric correlations exist between a number of parameters in all of those
functionals. For example, linear parametric correlations exist between the parameters $g_2$ and
$g_3$ which are responsible for the density dependence in the isoscalar channel
of the NLME model \cite{AAT.19}.
For the DDME functionals, the parameters $b_{\sigma}$ and $c_{\omega}$ vary linearly with
$c_{\sigma}$. Similarly, linear correlations are visible in the parameter pairs ($\alpha_V$,$\alpha_S$),
($\delta_V$,$\delta_S$), and ($\gamma_V$,$\gamma_S$) of the  PC-X functionals.
The observation of correlations effectively reduces the number of independent parameters to five
or six dependent on the structure and the underlying functional. In particular, the difference between 
the number of independent parameters depends on whether there is a density dependence in the 
isovector channel.
Thus, these numbers represent a limit of how many independent parameters could be defined in
the CDFT using fitting protocols based on ground state and nuclear matter properties. Of course,
at this stage, we cannot confirm that these correlations will also show up also for other
fitting protocols, in particular, for those containing other types of data. However, the presently
obtained results seem to be rather general.

  It is  reasonable to expect that similar parametric correlations also exist in
non-relativistic energy density functionals. In fact, in this case one should expect
even more such parametric correlations because as it is known a non-relativistic approximation
of covariant functionals in terms of a $p/M$-expansion leads to a non-relativistic functionals with
a large number of terms~\cite{Thies1985_PLB162-255,Thies1986_PLB166-23,PhD_Koenig1996}.
However, the various parameters in such functionals are not independent, but determined by
Lorentz invariance. An example are the Galilean invariant terms in some Skyrme
functionals~\cite{EBGKV.75,DD.95} connecting time-even and time-odd components of the functionals.
They are a direct consequence of the fact that time-even and the time-odd components in relativistic
 functionals are determined by the same coupling constants.

\section{ACKNOWLEDGMENTS}
The material is based upon work supported by the U.S. Department of Energy,
Office of Science, Office of Nuclear Physics under Award No. DE-SC0013037,
by the DFG cluster of excellence \textquotedblleft
Origins\textquotedblright\ (www.origins-cluster.de) and by
Ghana Atomic Energy Commission, National Nuclear Research Institute, Ghana.

\bibliography{references-24.bib}
\end{document}